\documentclass[modern]{aastex63}

\usepackage{natbib}
\usepackage{graphics,graphicx,mhchem,amssymb,commath}
\usepackage{appendix}
\bibpunct{(}{)}{;}{a}{,}{,}

\newcommand{\GG}[1]{}
\usepackage[symbol]{footmisc}

\begin{document}

\title{Detection of CH$_2$NH on Titan}

\correspondingauthor{Alexander E. Thelen}
\email{athelen@caltech.edu}

\renewcommand{\thefootnote}{\fnsymbol{footnote}}
\setcounter{footnote}{1}

\author{Alexander E. Thelen}
\affiliation{Division of Geological and Planetary Sciences, California
  Institute of Technology, Pasadena, CA, USA}
\affiliation{Solar System Exploration Division, NASA Goddard
  Space Flight Center, Greenbelt, MD, USA}

\author{Conor A. Nixon}
\affiliation{Solar System Exploration Division, NASA Goddard
  Space Flight Center, Greenbelt, MD, USA}

\author{Martin A. Cordiner}
\affiliation{Solar System Exploration Division, NASA Goddard
  Space Flight Center, Greenbelt, MD, USA}
\affiliation{Department of Physics, Catholic University
    of America, Washington DC, USA}

\author{Steven B. Charnley}
\affiliation{Solar System Exploration Division, NASA Goddard
  Space Flight Center, Greenbelt, MD, USA}

\author{V{\'e}ronique Vuitton}
\affiliation{Universit{\'e} de Grenoble Alpes, CNRS, IPAG, Grenoble, France}

\author{Benjamin Benne}
\affiliation{Laboratoire d'Astrophysique de
      Bordeaux, Universit{\'e} de Bordeaux/CNRS, Pessac, France}

\author{Jean-Christophe Loison}
\affiliation{Institut des Sciences Mol{\'e}culaires, Universit{\'e} de
  Bordeaux, Talence, France}

\author{Kevin M. Hickson}
\affiliation{Institut des Sciences Mol{\'e}culaires, Universit{\'e} de
  Bordeaux, Talence, France}

\author{Nicholas A. Teanby}
\affiliation{School of Earth Sciences, University of Bristol, Bristol, United Kingdom}

\author{Katherine de Kleer}
\affiliation{Division of Geological and Planetary Sciences, California
  Institute of Technology, Pasadena, CA, USA}

\author{Bryan Butler}
\affiliation{National Radio Astronomy Observatory, Socorro, NM, USA}

\author{Zbigniew Kisiel}
\affiliation{Institute of Physics, Polish Academy of Sciences, Warszawa, Poland}

\author{Karen Willacy}
\affiliation{Jet Propulsion Laboratory, California Institute of Technology,
  Pasadena, CA, USA}

\author{Michel Dobrijevic}
\affiliation{Laboratoire d'Astrophysique de
      Bordeaux, Universit{\'e} de Bordeaux/CNRS, Pessac, France}

\author{Patrick G. J. Irwin}
\affiliation{Department of Physics, University of Oxford, Oxford, United Kingdom}

\begin{abstract}
  
Saturn's moon, Titan, presents a chemical
environment where abiotic synthesis of organics essential for life
-- including amino acids and nucleotide bases -- may naturally occur.
The assessment of Titan's organic content begins with characterizing its atmospheric chemical
inventory, where many predicted molecular species remain unidentified. Here, we present the definitive
spectroscopic detection of \ce{CH2NH} (methanimine) on Titan from
observations with the Atacama Large Millimeter/submillimeter
Array. \ce{CH2NH} represents the first stable imine (\ce{C=N}) to be
detected on Titan, and may be integral for further biomolecule formation as a precursor
to glycine and adenine. We find
abundances of \ce{CH2NH} at the 0.1--1 ppb level in Titan's mesosphere
through multiple valid vertical profile configurations, yielding total
column densities ranging from $\sim0.9$--$3\times10^{13}$
  cm$^{-2}$. Middle-atmospheric abundances are commensurate with recent
photochemical model predictions. However, models with parameterized galactic
cosmic ray (GCR) fluxes, which exhibit a secondary, stratospheric formation site, 
are shown to provide poor fits to the data. Thus, our results provide
confirmation of predicted formation and destruction pathways for \ce{CH2NH} in the upper atmosphere, 
and demonstrate secondary formation mechanisms in the lower atmosphere require
adjustments. The latter presents an important barrier for fully estimating
Titan's lower atmospheric composition, as enhanced production by GCR is predicted for 
additional N-bearing species, such as \ce{NH3} and \ce{CH3NH2} (methylamine). The detection of \ce{CH2NH} provides the
motivation to search for additional species fundamental for prebiotic
molecule formation, which will
dramatically improve our understanding of complex molecular synthesis
of relevance to astrobiology in \ce{CH4}-rich environments.
  
\end{abstract}

\section{Introduction} \label{sec:intro}
\renewcommand{\thefootnote}{\arabic{footnote}}
\setcounter{footnote}{0}

Titan harbors a unique, chemically-diverse 
atmosphere maintained by the photodissociation of abundant molecular nitrogen (N$_2$) and
methane (CH$_4$). Photolysis products of these species interact
through neutral and ionic chemistry to synthesize an astounding array
of organic compounds \citep{nixon_24}; as such, Titan's known chemical inventory 
exceeds that of its parent body, Saturn, and the other giant
planets. Secondary to solar UV photons, Titan's innate atmospheric
chemistry is also driven out of equilibrium 
by charged particles from the 
Saturnian magnetosphere \citep{cravens_05, cravens_08} and intersystem
chemical re-distribution (e.g., from Enceladus, which may be a primary
source of Titan's diminutive oxygen content; \citealp{horst_08}), while dynamical and
meteorological processes occurring in Titan's middle and lower 
atmosphere govern the horizontal, vertical, and seasonally variable 
composition (see, e.g., \citealp{teanby_08b, teanby_09b,
  teanby_10a,vinatier_15}). The influence of both dynamical and chemical
complexity are evident on Titan's varied surface terrain (including
its hydrocarbon lakes; \citealp{turtle_11, neish_15, lopes_20,
  solomonidou_20}), though to what extent is not fully
understood. Further still, the potential for organics to make their
way into Titan's purported (though contested) sub-surface ocean
\citep{sotin_21, petricca_25} makes Titan an object of
immense astrobiological interest, though both the mechanisms and magnitude by which the
aqueous chemistry is ultimately subject to the atmospheric products
are currently intangible \citep{lopes_26}.

Remote-sensing measurements from both spacecraft and ground-based
facilities have made strides to characterize Titan's atmospheric
composition and its seasonal variability. The initial detections of
rudimentary cyanides and nitriles (i.e., compounds containing a carbon-nitrogen triple bond,
\ce{-C#N}) -- including HCN \citep{hanel_81}, C$_2$N$_2$, 
HC$_3$N \citep{kunde_81}, and CH$_3$CN \citep{marten_02} -- with
\textit{Voyager 1} and millimeter-wave facilities prompted further 
assessment of Titan's chemical complexity \citep{yung_84,
  cerceau_85, raulin_02}. The Cassini/Huygens mission, whereupon the majority of
our understanding of Titan is now predicated, revealed positive and
negative ionospheric compounds with masses (m/z) ranging from 100s to
1000s with the Ion and Neutral
Mass Spectrometer (INMS; \citealp{waite_05})
and Cassini Plasma Spectrometer (CAPS; \citealp{coates_07})
instruments, respectively. Many of these compounds were shown to
persist down to the lower atmosphere and surface from mass
spectrometer measurements onboard the Huygens probe
\citep{niemann_05}. These data
provided myriad potential chemical candidates and indicated plausible pathways to larger
organics \citep{vuitton_06, vuitton_07}, though definitive
identification of many trace species was (and remains) impossible
through those mass spectroscopic data alone due to limitations in
instrument resolution and ambiguities caused by degenerate masses
amongst molecular
species and isomers. Among the presumed chemical products
are molecules important for prebiotic chemistry, 
such as NH$_3$ (ammonia), CH$_2$NH (methanimine; also methylene
imine), CH$_3$NH$_2$ (methylamine), inferred to exist by
their protonated counterparts (e.g., NH$_4^+$, CH$_2$NH$_2^+$,
CH$_3$NH$_3^+$; 
\citealp{vuitton_06, vuitton_07, cui_09a, cui_09b, yelle_10}).

With the exceptions of NH$_3$ (heretofore only measured through NH$_4^+$)
and HNC (which is susceptible to rapid 
  conversion to HCN in Titan's upper atmosphere; \citealp{moreno_11}), Titan's confirmed N-bearing
  neutral gases all contain CN functional groups. Constraining the abundance of imines
  (\ce{C=N}), such as CH$_2$NH, on Titan informs a crucial branch of nitrogen
  chemistry enabled by a double bond, which
increases the availability of nitrogen in associated chemical reactions. While detected in astrophysical
sources beyond the Solar System (see \citealp{suzuki_16}, \citealp{mcguire_22}, and
references therein), gas-phase CH$_2$NH and related molecular
  species have yet to be detected
in extraterrestrial atmospheres. Laboratory studies on the
hydrogenation of Titan aerosol analogs -- ``tholins''
\citep{sagan_79} -- demonstrate the possibility of prebiotic molecule
formation on Titan under a variety of environmental conditions and
compositions \citep{neish_10, horst_12, poch_12}. Viable abiotic
synthesis sites include impact
craters, where increased heat and processing improve reaction rates
of Titan's organics with liquid water \citep{neish_18}. However, amino acids and nucleobases 
were not able to be confirmed by Cassini mass spectroscopy
instruments, and the surface conditions of sites of interest for
astrobiology may not be fully characterized until the landing of the
\textit{Dragonfly} mission.

The definitive detection of CH$_2$NH on Titan reinforces routes to
prebiotic molecules such as the amino acid glycine (\ce{NH2CH2CO2H}),
and the nucleotide base adenine (\ce{C5H5N5}), through its
links to both smaller (e.g., HCN, NH$_3$) and larger (e.g.,
CH$_3$NH$_2$, \ce{NH2CH2CN}) molecules of astrobiological importance
\citep{lavvas_08a, loison_15, suzuki_16, vuitton_19}. Thus, the
detection of CH$_2$NH and similar molecular species in Titan's
atmosphere vastly
improves our ability to assess Titan's potential for habitability until samples from the surface can be directly analyzed. 
Definitive identifications of Titan's trace gas species previously inferred from
their ionic counterparts have been achieved with recent observations
from the Atacama Large Millimeter/submillimeter Array (ALMA; 
\citealp{cordiner_15, palmer_17, nixon_20, thelen_20}). Here, we present
the detection and abundance estimate of CH$_2$NH -- the first stable
imine found on Titan -- through high sensitivity ALMA observations acquired
during 2024, 2025, and 2026 at $\sim1.22$ mm (245 GHz) and
$\sim0.944$ mm (317 GHz).

\section{ALMA Observations}\label{sec:obs}

We acquired ALMA observations targeting new chemical species in
Titan's atmosphere that were executed during ALMA Cycle 11 in 2024 December 
and 2025 January (ALMA Project Code $\#$ 2024.1.00570.S; denoted here
as P1). These observations were split into two spectral
settings in ALMA Band 6 (211--275 GHz; $\sim1.1$--1.4 mm): one
targeting the CO $J=2\rightarrow1$ transition at 230.538 GHz,
and a second at frequencies near $\sim$247 GHz targeting many 
transitions of various trace gas species. To improve the potential for
detecting weak lines, the observations were designed to achieve 
angular resolutions similar to that of Titan's angular scale ($\sim0.8$--$1''$) on the
sky. Following the detection of the CH$_2$NH $J=4_{1,4}\rightarrow3_{1,3}$
transition at 245.126 GHz in P1,
follow-up ALMA observations were approved under the Cycle 12
Director's Discretionary Time allocation (ALMA Project Code $\#$
2025.A.00010.S; hereafter P2). This program was dedicated to detecting strong
transitions of CH$_2$NH in ALMA Band 7 (275--373 GHz; $\sim0.8$--1.1
mm), and was executed in 2025 December and 2026
April. The observing
parameters for both programs are summarized in Table \ref{tab_obs}.

\begin{deluxetable*}{lllllll}
  \tablecaption{ALMA Observational Parameters\label{tab_obs}}
  \tablecolumns{7}
   \tablehead{Identifier & Project & Date & Freq. & Beam & Int. & $\#$ \\
     & Code & & (GHz)$^a$ & Size ($''$)$^b$ & Time (s) & Ant. \\ [-2.5ex]}
   \startdata
   P1 & 2024.1.00570.S & 2024 Dec. 15 & 230.5 & $1.57\times0.75$ & 787 & 45 \\
   P1 & 2024.1.00570.S & 2024 Dec. 31 & 247.5 & $0.99\times0.78$ & 2994 & 47 \\
        &  & 2025 Jan. 01 & & & 2996 & 46 \\
   P2 & 2025.A.00010.S & 2025 Dec. 22 & 317.4 & $0.47\times0.40$ & 2904 & 48\\
        &  & 2026 Apr. 21 & & & 2905 & 41 \\
        & & 2026 Apr. 26 & & & 5804 & 42 \\
  \enddata     
 \footnotesize
 \tablecomments{$^a$Additional spectral windows surrounding each
   central frequency were obtained (within 4 GHz of the listed
   frequencies). The full details on the 
   frequency range and resolution of each spectral window, along with the
   original data, can be found
   \href{https://almascience.nrao.edu/aq/?observationsSourceName=Titan&observationsProjectCode=2024.1.00570}{here
     for 2024} and
   \href{https://almascience.nrao.edu/aq/?observationsSourceName=Titan&observationsProjectCode=2025.A.00010.S}{here
     for 2026}. $^b$The full-width at half-maximum of the synthesized
   ALMA beam following deconvolution, analogous to the point-spread-function.}
\end{deluxetable*}

We calibrated both sets of ALMA observations using the National Radio Astronomy
Observatory ALMA Pipeline in the Common Astronomy Software
Applications (CASA; \citealp{jaeger_08}) package version 6.6.6. To aid in bandpass, phase, and
amplitude calibrations, quasars with known continuum properties at
millimeter wavelengths were observed alongside Titan
in both Band 6 (J2246--1206, J2301--0158) and Band
7 (J0006--0623, J0017--0512, J0051--0650) projects. Spectral windows
were set to 244, 488, or 976 kHz spectral resolution across both programs; CH$_2$NH spectra
were measured at a resolution of 488 kHz in P1 and 244 kHz in P2. We made minor
modifications to the standard ALMA pipeline procedures to increase the bandpass
smoothing intervals (to 10--30 channels depending on spectral
resolution; see \citealp{thelen_20} for details) during bandpass
calibration, which provided modest SNR
improvements in the final spectra.

We performed deconvolution on the resulting calibrated data using the
\texttt{tclean} function in CASA to produce spectral image cubes. To
reduce dilution of the disk-averaged data across large beam areas, we
applied the Briggs baseline weighting scheme \citep{briggs_95} with a
Robust value = 0.5, which resulted in
synthesized beam sizes comparable to Titan itself or slightly smaller
(see Table \ref{tab_obs}). The cell (pixel) sizes were set to $0.05''$, and iterative
\texttt{clean} cycles were performed until a threshold $\sim2\times$ the
spectral noise ($\sim2$ mJy in P1, 5--10 mJy in P2) was
achieved. Due to the relatively large angular size of the ALMA
synthesized beam compared to Titan, these observations were not
particularly well-suited to mapping the spatial distribution of the
detected gas species; the resulting CH$_2$NH
distributions are not indicative of enhancement at
  high latitudes in either observing
program (see Appendix \ref{app:obs}).

\begin{figure}[h]
  \centering
  \includegraphics[width=0.95\textwidth]{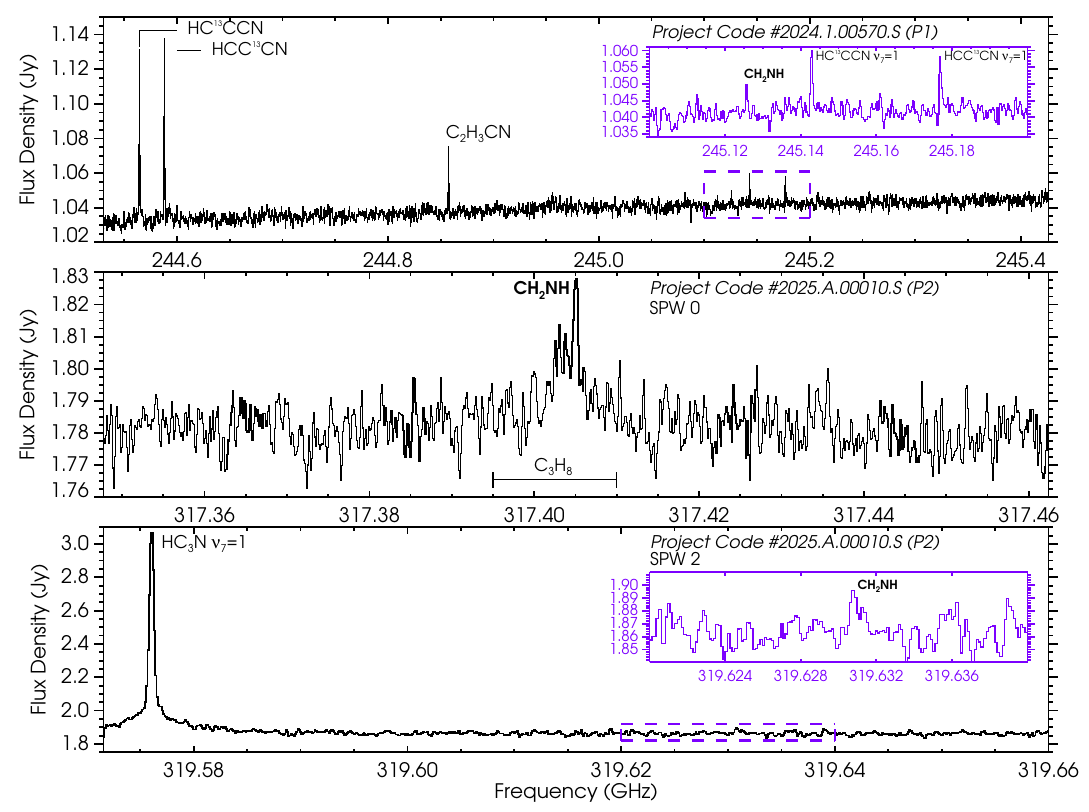} %{../Figures/ch2nh_da_spec}
  \caption{Full disk-averaged spectra of Titan from observations in
    2024--2025 (P1; top
    panel) and in two spectral windows in data from 2025--2026
    (P2; middle and bottom panels). Major detected gas species
    are labeled, including transitions of CH$_2$NH (bold). Insets
    in top and bottom panels (purple) show CH$_2$NH regions.}\label{fig:da_spec}
\end{figure}

The calibrated disk-averaged ALMA spectra containing CH$_2$NH emission are shown
in Figure \ref{fig:da_spec}. The CH$_2$NH
$J=4_{1,4}\rightarrow3_{1,3}$ transition in P1 (Figure
\ref{fig:da_spec}, top panel) was measured at a peak channel (i.e., the maximum
line flux density measured above the continuum) significance =
4.7-$\sigma$. Two transitions were detected in separate spectral
windows (SPW) in data from P2: CH$_2$NH $J=5_{0,5}\rightarrow4_{0,4}$
in SPW0 at a peak channel significance level of 5.8-$\sigma$, and CH$_2$NH
$J=5_{2,4}\rightarrow4_{2,3}$ in SPW2 at 3.3-$\sigma$
(Figure \ref{fig:da_spec} middle and bottom panels,
respectively). We also detected transitions of other well-known nitrile (e.g.,
HC$_3$N) and hydrocarbon (C$_3$H$_8$) species across these and other
spectral windows in both P1 and P2 observations -- a full list of rotational
transitions detected and analyzed in this study is provided in Appendix \ref{app:obs}.

\section{Radiative Transfer Modeling and Results}\label{sec:rt}

Following data calibration and imaging, we calculated the array of
emission angles corresponding to the disk-averaged spectra (Figure
\ref{fig:da_spec}) as
in previous works utilizing ALMA data (see, e.g.,
\citealp{thelen_18, thelen_19a}); the generation of sufficient emission
angles for disk-averaged spectral modeling -- including Titan's extended limb
emission -- is inherited from \citet{teanby_13}. We then performed radiative transfer
calculations to determine the temperature structure and composition that best
represented the state of Titan's atmosphere during this period through
the comparisons of synthetic model spectra to the data. We used the
Non-linear optimal Estimator for Multi-variatE spectral analySIS
(NEMESIS; \citealp{irwin_08}) software package for the spectral
inversion and iterative $\chi^2$-minimization in ``line-by-line'' mode, with input parameters
from previous works, laboratory and computational measurements where
available. We obtained spectral line data for CO, HC$_3$N (including its
vibrational modes and isotopologues), and CH$_2$NH 
from the Cologne Database for Molecular
Spectroscopy\footnote{See individual gas entries, including spectral
  line data and the corresponding computational
  references, here: https://cdms.astro.uni-koeln.de/classic/entries/ .}
(CDMS; \citealp{krause_89, muller_05, dore_12, endres_16}); we retrieved C$_3$H$_8$ spectral
line data from the Jet Propulsion Laboratory Molecular Spectroscopy
Database\footnote{https://spec.jpl.nasa.gov/ftp/pub/catalog/catdir.html},
and updated C$_2$H$_5$CN parameters were taken from \citet{kisiel_20}. We
fit the partition function for each molecular species as a function of
temperature using coefficients from the
corresponding spectroscopy databases. We adopted the temperature broadening and
full-width half-maximum parameters for all species as in previous
studies using ALMA and Cassini infrared observations
\citep{vinatier_07a, cordiner_15, serigano_16, molter_16, teanby_18,
  thelen_18}, and assumed $\Gamma=0.075$ cm$^{-1}$ and $n=0.5$ for
CH$_2$NH as in \citet{teanby_18} by analogy to C$_2$H$_5$CN spectral
line parameters.

We initialized atmospheric models of Titan using N$_2$, CH$_4$, and
H$_2$ gas profiles measured by the Huygens lander \citep{niemann_05,
  niemann_10}. While these gas species do not emit strongly in the
millimeter, collisionally-induced absorption (CIA) from these
molecular species influences the continuum emission; CIA parameters from self
and paired combinations of N$_2$, CH$_4$, and H$_2$ were assumed from
the works of \citet{borysow_86a, borysow_86b, borysow_86c,
  borysow_87}, \citet{borysow_91}, and \citet{borysow_93}. We obtained Titan's tropospheric
temperatures from Cassini Radio Science measurements from
\citet{schinder_20}, while the temperature profile from the
stratosphere to the mesosphere was initialized with the disk-averaged
profile from \citet{thelen_24b} measured with ALMA in 2022. We
compared modeled
continuum emission to the calibrated ALMA data to determine spectral
flux density offsets caused by minor ALMA flux calibration uncertainties (on order
$\sim5$--$10\%$ in ALMA Band 6 and 7) and averaging of the
data over the relatively large ALMA beam area \citep{thelen_18}. We adjusted for these
offsets by applying multiplicative scaling factors to each
spectrum.

\subsection{Temperature and Gas Retrievals}
\begin{figure}[h]
 \centering
 \includegraphics[width=0.95\textwidth]{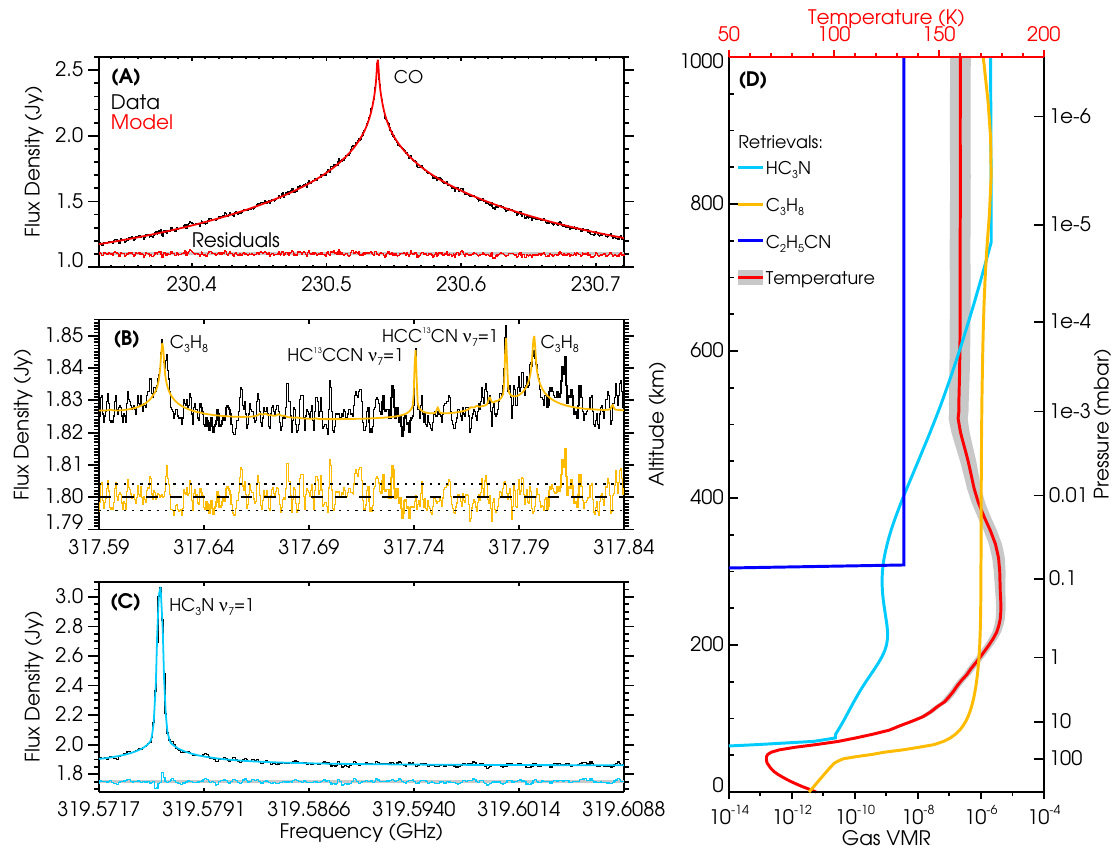} %{../Figures/spec_temp_gas}
 \caption{Retrieval of Titan's temperature profile and gas abundances for additional
   detected molecular species. (A) Disk-averaged spectrum of the CO $J=2\rightarrow1$
   transition from ALMA observations in P1 (black). The best-fit model
 spectrum generated by NEMESIS is shown (red) following the retrieval
 of a continuous temperature profile. The corresponding residual
 (data-model) spectrum is shown below, offset by +1.1 Jy for
 visibility. The 1-$\sigma$ noise level is shown with the 
 residual spectrum (gray envelope). (B) ALMA data from P2/SPW1 (black)
    compared to a best-fit model of C$_3$H$_8$ and HC$_3$N
    isotopologue emission (gold spectrum). The residual spectrum is offset by +1.8 Jy for visibility. The
    1-$\sigma$ noise level is shown bracketing the residual spectrum (dotted lines). (C)
    HC$_3$N $\nu_7=1$ spectrum from P2/SPW2 (black) with best-fit
    model produced through a continuous retrieval (cyan). The residual
    spectrum is adjusted for clarity (+1.75 Jy), and the
    1-$\sigma$ level is denoted by the gray envelope. (D) The corresponding retrieved
    vertical gas and temperature profiles from panels A, B, and C. The
    retrieved temperature profile from panel A is shown in red, with
    retrieval errors denoted by the gray error envelope. The vertical structure of
    C$_3$H$_8$ was initialized from the model of
    \citet{vuitton_19}. The retrieved C$_2$H$_5$CN profile is also
    shown, retrieved from P2/SPW0.}\label{fig:combo_1}
\end{figure}

We determined Titan's disk-averaged
temperature profile by performing a retrieval on the CO
$J=2\rightarrow1$ spectrum from P1 (Figure \ref{fig:combo_1}A). We
maintained the tropospheric temperature profile from Cassini and held
the CO volume mixing ratio (VMR) at 50 ppm \citep{serigano_16, thelen_18} and allowed the
temperature profile to vary continuously at all atmospheric layers
above the tropopause. Temperature profile errors were initially set to
1--5 K as a function of altitude, and were further constrained at altitudes
where the data are more sensitive to variations in atmospheric
temperature ($\sim100$--450 km; \citealp{thelen_18}).
Above $\sim$500 km where the sensitivity of the CO $J=2\rightarrow1$
spectral line to thermal variability is low \citep{serigano_16, thelen_18}, the retrieved temperature profile relaxed to the input
isothermal profile at 160 K. The resulting temperature profile is
shown in Figure \ref{fig:combo_1}D.

After retrieving the latitudinally averaged temperature profile from
the CO spectrum, we fit spectra from P1 and P2 to retrieve individual
gas profiles for C$_3$H$_8$, C$_2$H$_5$CN, HC$_3$N, and finally
CH$_2$NH. Input profiles for C$_2$H$_5$CN and HC$_3$N 
  were obtained from recent studies using ALMA \citep{lai_17,
    thelen_19a, nixon_20}, while the C$_3$H$_8$ abundance profile was
  assumed from photochemical model predictions (\citealp{dobrijevic_16}\footnote{The full set of nominal profiles from
    \citet{dobrijevic_16} are available
    \href{https://perso.astrophy.u-bordeaux.fr/~mdobrijevic/photochemistry/titan_model1.php.html}{here}.},
  \citealp{vuitton_19}). The HC$_3$N $\nu_7=1$ transition in P2
(Figure \ref{fig:da_spec}, bottom panel) was bright and well-resolved such that
we were able to perform a continuous retrieval of the HC$_3$N
abundance as a function of altitude, similar
to the temperature retrieval for CO. The input profiles for
C$_2$H$_5$CN and C$_3$H$_8$ were scaled to fit the corresponding
spectra. The resulting best-fit HC$_3$N $\nu_7=1$ and C$_3$H$_8$ spectra are
  shown in Figure \ref{fig:combo_1}, along with the retrieved abundance
  profiles. Finally, we used the retrieved temperature
and vertical C$_3$H$_8$, HC$_3$N profiles to perform fits to the CH$_2$NH spectra in P1 and P2
separately to retrieve CH$_2$NH abundance profiles and column
densities. The subtraction non-CH$_2$NH gas emission in the P2/SPW0
spectrum (Figure \ref{fig:da_spec}, middle panel), specifically, is detailed in
Appendix \ref{sec:decomp}. 

\subsection{CH$_2$NH Retrievals and Results}
As the CH$_2$NH emission lines we observed in both projects are narrow
and fairly
low signal-to-noise ($\sim3$--6-$\sigma$), good fits (i.e., $\chi^2/n\approx1$)
were achievable by scaling input model profiles as opposed to
performing continuously varying retrievals. Neither the total
column density nor vertical VMR structure of CH$_2$NH are known in Titan's atmosphere, so we tested multiple input
CH$_2$NH abundance profiles. These include simple parameterizations
(e.g., constant above a certain altitude, or increasing linearly in
pressure space -- ``step'' and ``gradient'' profiles,
respectively) and profiles generated through current photochemical
models that incorporate multiple chemical sources and
sinks from hydrocarbon and nitrogen chemistry 
\citep{willacy_16, dobrijevic_16, vuitton_19}. The photochemical simulations are
parameterized differently, are derived from both known and approximated
chemical kinetics, and include tunable external energetics (e.g.,
magnetospheric electrons, GCR); the comparison of fits to the data
between models thus allows for the assessment of the included
formation and destruction pathways for CH$_2$NH throughout the
atmosphere. 

\begin{figure}[h]
  \centering
  \includegraphics[width=0.95\textwidth]{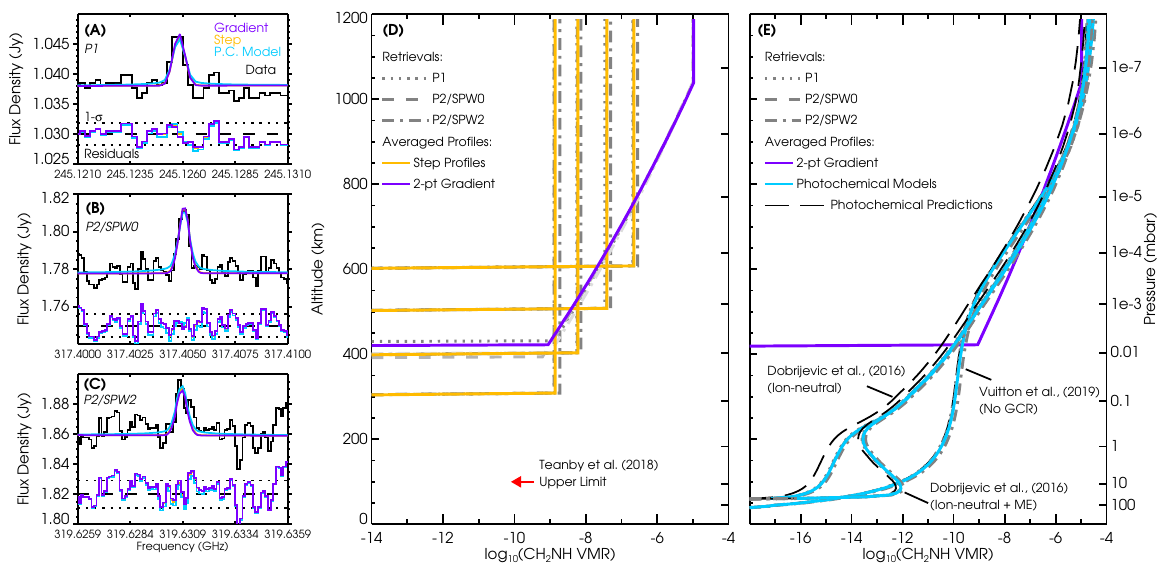} %{../Figures/spec_vert_comp}
  \caption{Best-fit model spectra and corresponding vertical
    profiles. (A) Disk-averaged spectrum (black) of CH$_2$NH
    $J=4_{1,4}\rightarrow3_{1,3}$ feature
    in P1 following the subtraction of other gases and compared to best-fit model
    spectra produced by representative gradient (purple), step (gold),
    and photochemical (P.C.; cyan) profiles. The corresponding
    residual (data-model) spectrum is shown below, offset by +1.03 Jy for visibility (dashed
    line). 1-$\sigma$ uncertainties are denoted by dotted lines above
    and below the residual spectrum. (B) CH$_2$NH
    $J=5_{0,5}\rightarrow4_{0,4}$ spectrum from P2/SPW0 following the
    spectral decomposition as detailed in Appendix
    \ref{sec:decomp}. The residual is offset by +1.75 Jy. (C) CH$_2$NH
    $J=5_{2,4}\rightarrow4_{2,3}$ spectrum from P2/SPW2. (D)
    Comparison of simple, best-fit vertical CH$_2$NH profiles
    parameterized in NEMESIS: step (constant VMR above a certain
    altitude) profiles at 300, 400, 500, and 600 km; a log-linear
    gradient (in pressure space) fit between two pressure-VMR
    points. Individual best-fit
    models for each spectrum are shown in gray (P1: dotted; P2/SPW0:
    dashed; P2/SPW2: dash-dot), with
    weighted averaged profiles shown in color (step profiles in gold;
    gradient in purple). The stratospheric (\textgreater50 km)
    upper limit
    calculated by \citet{teanby_18} is shown as a red arrow for
    reference. (E) Retrievals produced by scaling photochemical
    model predictions (long dashed lines). Individual (gray) and weighted
  averaged profiles (cyan) are shown for photochemical profiles that provide
  good fits to the data: the nominal (ion-neutral) models from
  \citet{dobrijevic_16} and \citet{vuitton_19}, and the
  \citet{dobrijevic_16} ion-neutral model with the addition of
  magnetospheric electron flux. The gradient from panel D is also shown for reference.} \label{fig:combo_2}
\end{figure}

Examples of the best-fit ($\chi^2/n\approx1$) spectra for all three detected CH$_2$NH transitions are shown in
Figure \ref{fig:combo_2}A--C, with the corresponding retrieved vertical profile for
each feature in Figure
\ref{fig:combo_2}D and E. Averaged profiles (colored lines in Figure
\ref{fig:combo_2}D,E) were calculated from an error-weighted average of
the three profiles resulting from the individual spectral fits, with
weights determined from the formal retrieval errors. We found that each input vertical profile resulted in consistent
retrieved VMRs when compared across the
three CH$_2$NH spectra. Multiple model parameterizations provided fits of similar statistical
significance and indistinguishable residual amplitudes such that there
exist an ensemble of vertical profiles that best fit the ALMA
data. This is not a unique situation; multiple sets of model profile
configurations were employed to characterize the abundance of
C$_2$H$_5$CN, C$_2$H$_3$CN, 
c-C$_3$H$_2$, and CH$_3$C$_3$N on Titan \citep{cordiner_15, palmer_17,
  nixon_20, thelen_20},
whose column densities, abundances and vertical structures were similarly
uncertain. 

The constant abundance profiles between 300--600 km 
and gradients anchored at similar altitude ranges increasing linearly
to the thermosphere (constrained by Cassini/INMS measurements;
\citealp{vuitton_06, vuitton_07}) all provided adequate fits to the
data. However, only the photochemical
models from \citet{dobrijevic_16}
and \citet{vuitton_19} without incorporated GCR chemistry in the lower atmosphere were
able to be scaled to fit the data. The model from
\citep{willacy_16} was able to provide an acceptable fit to the data,
although the resulting abundances in the upper
atmosphere are in excess of $10\times$ that measured by Cassini, and
the profile requires considerable scaling to make up for the lack of
predicted CH$_2$NH in the middle atmosphere. The full comparison of
photochemical models to the data are detailed in Appendix
\ref{sec:pc}.

When comparing the linear gradient profile to
scaled photochemical model profiles (Figure \ref{fig:combo_2}E), the CH$_2$NH profiles converge upon a common range of
abundances between $\sim$0.1--1 ppb that increase with altitude from the lower mesosphere (\textgreater400
km; \textless0.01 mbar) to the upper thermosphere ($\sim1000$ km) at
a VMR $\approx10$ ppm, as inferred by Cassini/INMS data 
\citep{vuitton_06, vuitton_07}. This is significantly
($\sim100$--$1000\times$) less than HCN, which has been observed
at abundances of $\sim1$ ppm in the middle atmosphere and $\gtrsim100$
ppm in the thermosphere (see \citealp{thelen_22}
and references therein). Other nitriles (e.g., HC$_3$N, CH$_3$CN)
range from  10--100$\times$ more abundant than our CH$_2$NH measurements in the mesosphere
\citep{thelen_19a}.

\begin{deluxetable}{llll}
  \tablecaption{CH$_2$NH Best-Fit Model Results \label{tab_abunds}}
  \tablecolumns{4}
    \tablehead{Model & VMR/Scaling & N (cm$^{-2}$)$^a$\\ [-2.5ex]}
    \startdata
    300 km Step & $(1.38\pm0.12)\times10^{-09}$ & $(1.95\pm0.17)\times10^{13}$ \\
400 km Step & $(5.98\pm0.52)\times10^{-09}$ & $(1.32\pm0.12)\times10^{13}$ \\
500 km Step & $(3.91\pm0.33)\times10^{-08}$ & $(1.12\pm0.10)\times10^{13}$ \\
600 km Step & $(2.22\pm0.19)\times10^{-07}$ & $(9.92\pm0.83)\times10^{12}$ \\
%\midrule
Gradient & & $(9.28\pm1.30)\times10^{12}$ \\
\phantom{spce} Point 1 (417.6 km) & $(8.01\pm1.29)\times10^{-10}$ & \\
\phantom{spce} Point 2 (1039 km) & $(1.05\pm0.13)\times10^{-05}$ & \\
%\midrule
Vuitton et al. (2019) [N] & $1.16\pm0.08$ & $(2.93\pm0.26)\times10^{13}$ \\
Dobrijevic et al. (2016) [N] & $3.32\pm0.22$ & $(9.09\pm0.79)\times10^{12}$ \\
Dobrijevic et al. (2016) [ME] & $1.56\pm0.11$ & $(1.19\pm0.10)\times10^{13}$ 
 \\
    \enddata
    \footnotesize
    \tablecomments{Retrieved abundances
    are presented for step and gradient models. The abundance and
    altitude values for the two points of the gradient model are listed separately. Scaling factor and error are
    shown in the case of scaled photochemical profiles. For photochemical profiles: [N] correspond to nominal
    models incorporating ion and neutral chemistry only; [ME]
    indicates the version of the \citet{dobrijevic_16} model including
    magnetospheric electrons. $^a$Total column density
    integrated up to 1200 km.}
\end{deluxetable}

Following the retrieval and averaging of best-fit
vertical profiles, we calculated the total CH$_2$NH column densities in
Titan's atmosphere up to 1200 km, which range 
from 0.91--2.93$\times10^{13}$ cm$^{-2}$. We note that the derived column densities
exhibit a relatively small spread despite
the variability in retrieved profile structures. These values, along with the
corresponding VMR and scaling factors for all valid vertical profiles, 
are listed in Table \ref{tab_abunds}.
Our column density measurements are similar to that calculated
by \citet{vuitton_19} = $4.1\times10^{13}$ when accounting for the
difference in integrated altitude levels (that study calculated column
density values
up to $\sim1500$ km, while our models extend to only 1200 km). The
column densities derived by \citet{cordiner_15} for C$_2$H$_5$CN (ethyl cyanide or propionitrile;
$\sim$1--5$\times10^{14}$ cm$^{-2}$) by and
\citet{palmer_17} for C$_2$H$_3$CN (vinyl cyanide or acrylonitrile;
$\sim0.4$--1.4$\times10^{14}$ cm$^{-2}$), are roughly 
an order of magnitude larger than derived for CH$_2$NH in this study. However, CH$_2$NH may be at similar or $\sim2\times$ the
column density of CH$_3$C$_3$N as found by \citet{thelen_20}. Thus, though
highly reactive, CH$_2$NH is not the least abundant gas 
detected so far with ALMA in Titan's middle atmosphere (the aforementioned CH$_3$C$_3$N and
similarly rarefied c-C$_3$H$_2$ found by \citealp{nixon_20}).

At altitudes \textless300 km (pressures $\gtrsim0.1$ mbar), our data are
not sufficiently sensitive to changes in CH$_2$NH to provide adequate
constraints (see Appendix \ref{sec:pc}); however, photochemical profiles produced by
\citet{dobrijevic_16} and \citet{vuitton_19} with the
inclusion of GCR chemistry produce significantly greater radiances
than the observed data due to a
secondary, stratospheric CH$_2$NH formation site resulting in
unfavorable fits to the data. This corroborates the findings of
\citet{teanby_18}, who calculated a stratospheric 
upper limit = 0.35 ppb -- abundances
greater than this produce CH$_2$NH spectra with broadened line wings
that cannot be reconciled by either the data analyzed in that study or
here (see Appendix \ref{sec:pc}). 

\section{Discussion}\label{sec:diss}
\subsection{CH$_2$NH in Titan's Atmosphere}
% Formation:
The confirmation of CH$_2$NH as a neutral product of Titan's
atmospheric photochemistry validates previous INMS data analyses
wherein m/z=30 counts were associated with protonated methanimine
(CH$_2$NH$_2^+$), prompting the inference of CH$_2$NH in the upper atmosphere at the
$\sim10$ ppm level \citep{vuitton_06, vuitton_07, cui_09a}. Other
candidate ion species were ruled out as being the primary contributor
to the measured INMS data at m/z=30 due to being highly 
reactive (C$_2$H$_6^+$) or produced at substantially lower rates
(NO$^+$), thus leaving CH$_2$NH$_2^+$ as the favored, though
unconfirmed, candidate \citep{vuitton_06, vuitton_07, yelle_10}. Ion
and neutral production of CH$_2$NH$_2^+$ and CH$_2$NH were factored
into subsequent photochemical models (e.g., \citealp{lavvas_08a,
  krasnopolsky_09, yelle_10, loison_15}), where
CH$_2$NH was predicted to form through a number of combinations with N$_2$
and CH$_4$ dissociation products and other N-bearing radicals, primarily:
\begin{eqnarray}
  \ce{N(^2D) + CH4 -> CH2NH + H} \label{ch2nh_f1} \\
  \ce{NH + CH3 -> CH2NH + H} \label{ch2nh_f2} 
\end{eqnarray}
and also through electron recombination with CH$_2$NH$_2^+$
\citep{cui_09a, yelle_10, loison_15}. Additional reactions
with other radicals, hydrocarbons, and NH$_3$ have been indicated as
potential secondary contribution sources at much lower production
rates:
\begin{eqnarray}
  \ce{2 H2CN -> HCN + CH2NH} \label{ch2nh_f3} \\
  \ce{N(^2D) + C2H6 -> CH2NH + CH3} \label{ch2nh_f4} \\
  \ce{NH3 + CH -> CH2NH + H} \label{ch2nh_f5}
\end{eqnarray}
(see \citealp{lavvas_08a} and \citealp{loison_15}, and references
therein). These reactions produce CH$_2$NH in Titan's upper atmosphere
($\gtrsim1000$ km), with a potential secondary stratospheric formation
site attributed to the dissociation of N$_2$ by GCR \citep{loison_15,
  vuitton_19}; as demonstrated by our retrievals to the ALMA
observations, however, the formation of CH$_2$NH through GCR chemistry
is over-predicted in the current
models (see Appendix \ref{sec:pc}). Previously, \citet{teanby_09a} noted discrepancies in GCR chemistry through
the comparison of C$_2$N$_2$ measurements with photochemical models from
\citet{lavvas_08a} and \citet{wilson_04}, so the impact of GCR on multiple
N-bearing species remains both important and uncertain.

Our retrieval results show that the photochemical predictions from
ion-neutral chemistry models (with and without magnetospheric electron
precipitation) in the middle
and upper atmosphere provide valid fits to the data, indicating the
chemistry here is fairly well constrained. The nominal
model of \citet{vuitton_19} required little scaling (a factor of 1.16) to
fit the data, while the models (without GCR chemistry) from
\citet{dobrijevic_16} required additional scaling (factors of
$\sim1.6$--3.3; Table \ref{tab_abunds}). Scaling the model of
\citet{willacy_16} provided adequate fits to the data (see Appendix \ref{sec:pc}), though it required
a multiplicative factor of $\sim85$ due to the dearth of CH$_2$NH produced at
altitudes \textless800 km. We note that all
retrieved profiles required positive scaling of their \textit{a
  priori} counterparts, indicating that there
may be more CH$_2$NH in the middle atmosphere than predicted by
models.

% Destruction
Chemical models predict loss of CH$_2$NH below its
upper atmospheric formation site due to photodissociation into other
products, e.g.:
\begin{eqnarray}
  \ce{CH2NH + h\nu{} -> HCN + H + H} \label{ch2nh_d1} \\ 
   \ce{-> H2CN + H}  \label{ch2nh_d2}
\end{eqnarray}

and through reactions with hydrogen:
\begin{eqnarray}
  \ce{H + CH2NH -> H2 + H2CN} \label{ch2nh_d3} \\
  \ce{H + CH2NH ->[M] CH2NH2} \label{ch2nh_d4} 
\end{eqnarray}
\citep{lavvas_08a, loison_15, vuitton_19}. Importantly, due to its
relatively high proton affinity, CH$_2$NH is efficiently recycled back
into CH$_2$NH$_2^+$ through combination with a variety of other
gas species \citep{yelle_10}. While electron
recombination of CH$_2$NH$_2^+$ can again produce CH$_2$NH, it also
acts as a sink through the formation of other trace products, including HCN and NH$_3$:
\begin{eqnarray}
  \ce{CH2NH2^+ + e^- -> HCN + H + H2} \label{ch2nh_d5} \\ 
  \ce{-> NH2 + CH2} \label{ch2nh_d6}
\end{eqnarray}
the latter reaction is important for the generation of amino
radicals (NH$_2$), which leads to ammonia production: 
\begin{eqnarray}
  \ce{NH2 + H2CN -> NH3 + HCN} \label{ch2nh_ex2}
\end{eqnarray}
\citep{yelle_10, vuitton_19}. The pathway to NH$_2$ and NH$_3$
formation places CH$_2$NH and CH$_2$NH$_2^+$ in a critical juncture of
Titan's atmospheric chemistry, as these products enable routes to
further complex molecules predicted to exist on Titan relevant for
prebiotic chemistry (e.g., \ce{NH2 + CH3 +M -> CH3NH2 + M}; \citealp{lavvas_08a}). 

\subsection{Towards Further Abiotic Chemistry}
% Comparisons to detections in other sources:
Spectroscopic measurements of complex molecular species in the interstellar medium
(ISM) -- including large N-bearing
molecules, alcohols, imines, and amines (\ce{X-NH2}) -- have
out-paced Solar System remote-sensing detections through ongoing ground-based
observing campaigns at long wavelengths (see the thorough census of \citealp{mcguire_22} and
references therein). These (relatively) low-temperature environments
are chemically modified by stellar UV photolysis, GCR flux, and conducive to ice
grain formation and subsequent chemistry, resulting in the postulated
abiotic formation
of biomolecules such as amino acids and nucleotide bases. Though multiple amino acids have been
detected in meteoritic, comet, and asteroid samples
\citep{kvenvolden_70, altwegg_16,
  glavin_25}, amino acids have yet to be definitively detected
in the ISM \citep{belloche_08}. CH$_2$NH and CH$_3$NH$_2$, however, have both
been detected towards high mass star forming regions -- in particular, the Sagittarius B2 (Sgr
B2) molecular cloud complex and the Kleinmann-Low Nebula in Orion
(Orion KL; see
review by \citealp{suzuki_16} and references therein). Aminoacetonitrile
(\ce{NH2CH2CN}) has also been detected toward Sgr B2
\citep{belloche_08}, and CH$_2$NH has even been detected in
extragalactic sources \citep{salter_08, muller_11}. These detections
indicate increased chemical complexity towards prebiotics in these
environments through gas and
ice phase chemistry.

Glycine is predicted to form in the ISM through hydrogenation of
CH$_2$NH into CH$_2$NH$_2$ or CH$_3$NH$_2$ \citep{kim_11, theule_11},
and from the former into \ce{NH2CH2CN} through combination with
HCN on grain surfaces \citep{danger_11}. Subsequently, either hydrolysis of \ce{NH2CH2CN} or combination of CH$_3$NH$_2$ with
CO$_2$ both lead to glycine \citep{peltzer_84, holtom_05,
  suzuki_16}. Additionally, as the photodissociation of CH$_2$NH and
electron recombination of CH$_2$NH$_2^+$ result in HCN (e.g., Equations
\ref{ch2nh_d1}, \ref{ch2nh_d5}), CH$_2$NH also provides an additional
constraint on pathways to adenine, produced through HCN polymerization
\citep{oro_61a, oro_61b, roy_07}. 

As opposed to a direct comparison of our retrieved CH$_2$NH abundances --
taken with respect to the bulk N$_2$ atmosphere on Titan -- to those
measured for other astrophysical sources (often calculated compared to H$_2$), we can
compare the ratio of CH$_2$NH to other common nitriles found in both
sources, such as HCN. The elevated production of CH$_2$NH and other
nitriles has been measured in warm ($\sim100$--300
K) sources, resulting in CH$_2$NH/HCN on order 0.01--0.1 \citep{dickens_97};
in cold ($\sim10$--50 K) sources, values closer to 0.001 have been
reported \citep{dickens_97, lin_26}. By comparison to the mesospheric
HCN profiles in \citet{thelen_22}, we find CH$_2$NH/HCN
$\approx1$--$5\times10^{-3}$ at 600 km, which is similar to cold
astrophysical sources such as the pre-stellar core L1544 and the outer
ridge of Orion KL (Ori 3N) \citep{dickens_97, spezzano_22, lin_26}. At lower altitudes on
  Titan, the combination of significant divergence between
  photochemical model predictions and the
  lack of sensitivity in our ALMA spectra preclude us from making
  constraints on the CH$_2$NH/HCN. Additional measurements of CH$_2$NH in
  the lower atmosphere and the assessment of GCR impacts on
  stratospheric chemistry may help to further elucidate the 
  possibility and rate of biomolecule synthesis among gas species and aerosols formed in Titan's stratosphere and troposphere.

Finally, though unresolved mass
spectra from Cassini/INMS cannot definitively
(dis)prove the existence of glycine on Titan \citep{horst_12}, both CO$_2$ and
H$_2$O exist on Titan in the gas phase
\citep{samuelson_83b, coustenis_98} and potentially on the surface
\citep{horst_17}. Combined with the confirmation of
CH$_2$NH, the ingredients for abiotic synthesis of amino acids exist on
Titan through the pathways proposed for the ISM. Indeed, \citet{horst_12} predicted the formation of
multiple amino acids and nucleotide bases under Titan-like conditions (though with enhanced CO
concentrations) in aerosol and gas-phase products through laboratory
experiments, as did \citet{neish_10} and \citet{poch_12} through hydrolysis of Titan tholins. Thus, though the chemistry remains somewhat poorly constrained for
complex organics, Titan presents a strong candidate for prebiotic
molecule formation through its unique interplay of atmospheric,
aerosol, and (sub)surface chemistry.

\section{Conclusion}\label{sec:conc}
The spectroscopic confirmation of CH$_2$NH here presents an important piece
of Titan's complex chemical composition through its chemical links to
NH$_3$, HCN and other N-bearing organics, and potentially biomolecules 
(e.g., glycine). The abiotic synthesis of amino acids and nucleotide
bases on Titan may be feasible through the
products of its atmospheric chemistry -- such as CH$_2$NH and
related molecules CH$_3$NH$_2$, \ce{NH2CH2CN} -- as shown in studies of
ice-grain chemistry for the characterization of the ISM. Additional laboratory and meteorological studies are
needed to assess potential for analogous formation pathways of glycine
and other prebiotic
species on ice-grains and haze conglomerates in Titan's stratosphere
and troposphere, and to quantify the precipitation rate of CH$_2$NH to commingle with CO$_2$ and H$_2$O
ices, organics, and
hydrocarbon lakes on Titan's diverse surface \citep{krasnopolsky_09,
  turtle_11, palmer_17}. The possible transfer of biologically relevant organics through
Titan's crust into the sub-surface
ocean or liquid reservoirs (e.g., through large impactors; \citealp{neish_24}) is 
tantalizing though not well understood. Further study of
the transfer and
stability of these molecules into Titan's sub-surface is
necessary 
to characterize the potential aqueous composition as influenced from above to 
estimate the formation of amino acids and other
biomolecules in the interior. While some of these uncertainties may be addressed
by the upcoming \textit{Dragonfly} mission, which will sample multiple
regions on Titan's surface in the 2030's \citep{barnes_21}, the knowledge of the CH$_2$NH abundance in Titan's
middle and upper atmosphere can be used in the short-term to further
inform our understanding of GCR chemistry by comparison with photochemical models for
the prediction of additional related species (e.g., amines
such as \ce{CH3NH2}, \ce{NH2CH2CN}) that may be observable at
millimeter-wavelengths.

\section{Acknowledgements}
This paper makes use of the following ALMA data:
ADS/JAO.ALMA$\#$2024.1.00570.S and $\#$2025.A.00010.S. ALMA is a
partnership of ESO (representing its member states), NSF (USA) and
NINS (Japan), together with NRC (Canada), NSTC and ASIAA (Taiwan), and
KASI (Republic of Korea), in cooperation with the Republic of
Chile. The Joint ALMA Observatory is operated by ESO, AUI/NRAO and
NAOJ. The National Radio Astronomy Observatory is a facility of the
National Science Foundation operated under cooperative agreement by Associated Universities, Inc.

Part of this research was carried out at the Jet Propulsion
Laboratory, California Institute of Technology.

The authors would like to acknowledge the following funding sources:
AET and KdK received funding from NASA grant $\#$80NSSC23M0092; MAC
was supported by NSF grant $\#$AST-2407709; NAT was funded by UK
Science and Technology Facilities Council grant ST/Y000676/1; KW was funded by NASA
grant $\#$80NM0018D0004 and $\#$25-SMDSS25-0037.

Finally, the authors would like to acknowledge A. Moullet and the staff of the NRAO
North American ALMA Science Center for their
support in the acquisition and reduction of these ALMA data.

\pagebreak

\begin{appendices}
  \appendix
  \section{Additional Observation Details and Images} \label{app:obs}
  \begin{deluxetable}{lllll}
    \tablecaption{Detected Spectral Transitions \label{tab_spec}}
    \tablecolumns{6}
   \tablehead{Gas & Rest & Transition$^a$ & $E'$ & Project \\
   Species & Freq. (GHz) & & (K) & /Spec. Win. \\ [-2.5ex]}
  \startdata
   CO & 230.538 & $2\rightarrow1$ & 16.60 & P1 \\
CH$_2$NH & 245.126 & $4_{1,4}\rightarrow3_{1,3}$ & 37.31 & P1  \\
CH$_2$NH & 317.405 & $5_{0,5}\rightarrow4_{0,4}$ & 45.85 & P2/SPW0 \\
CH$_2$NH & 319.631 & $5_{2,4}\rightarrow4_{2,3}$ & 77.56 & P2/SPW2 \\
HC$_3$N $\nu_7=1$ & 319.576 & $35_{-1}\rightarrow34_{1}$ & 597.0& P2/SPW2 \\
HC$^{13}$CCN & 244.564 & $27\rightarrow26$ & 164.3 & P1\\
HC$^{13}$CCN $\nu_7=1$ & 245.143 & $26_1\rightarrow25_{-1}$& 480.6 & P1\\
HC$^{13}$CCN $\nu_7=1$ & 317.739 & $35_1\rightarrow34_{-1}$& 590.4 & P2/SPW1 \\
HCC$^{13}$CN & 244.588 & $27\rightarrow26$ & 164.4 & P1\\
HCC$^{13}$CN $\nu_7=1$ & 245.177 & $27_1\rightarrow26_{-1}$& 483.1 & P1\\
HCC$^{13}$CN $\nu_7=1$ & 317.783 & $35_1\rightarrow34_{-1}$& 592.9 & P2/SPW1 \\
C$_2$H$_3$CN & 244.857 & $26_{2,25}\rightarrow25_{2,24}$ & 167.9 & P1\\
C$_2$H$_5$CN & 317.403 & $7_{6,1}\rightarrow6_{5,2}$ & 52.06 & P2/SPW0\\
C$_3$H$_8$ & 317.404 & $20_8\rightarrow20_7$& 225.8 & P2/SPW0 \\
C$_3$H$_8$ & 317.616 & $19_8\rightarrow19_7$ & 210.5 & P2/SPW1 \\
C$_3$H$_8$ & 317.796 & $18_8\rightarrow18_7$ & 195.9 & P2/SPW1
 \\
   \enddata
 \footnotesize
 \tablecomments{$^a$Rotational emission lines are denoted
   $J' \rightarrow J''$, $J'_{K'} \rightarrow J''_{K''}$, or
   $J'_{K_a',K_c'} \rightarrow J''_{K_a'',K_c''}$. Rows listed as
   C$_3$H$_8$ refer to spectral bands with representative frequencies
   and $J'_{K'} \rightarrow J''_{K''}$ notation, but encapsulate
   many transitions.}
 \end{deluxetable}

\begin{figure}[h]
 \centering
  \includegraphics[width=0.95\textwidth]{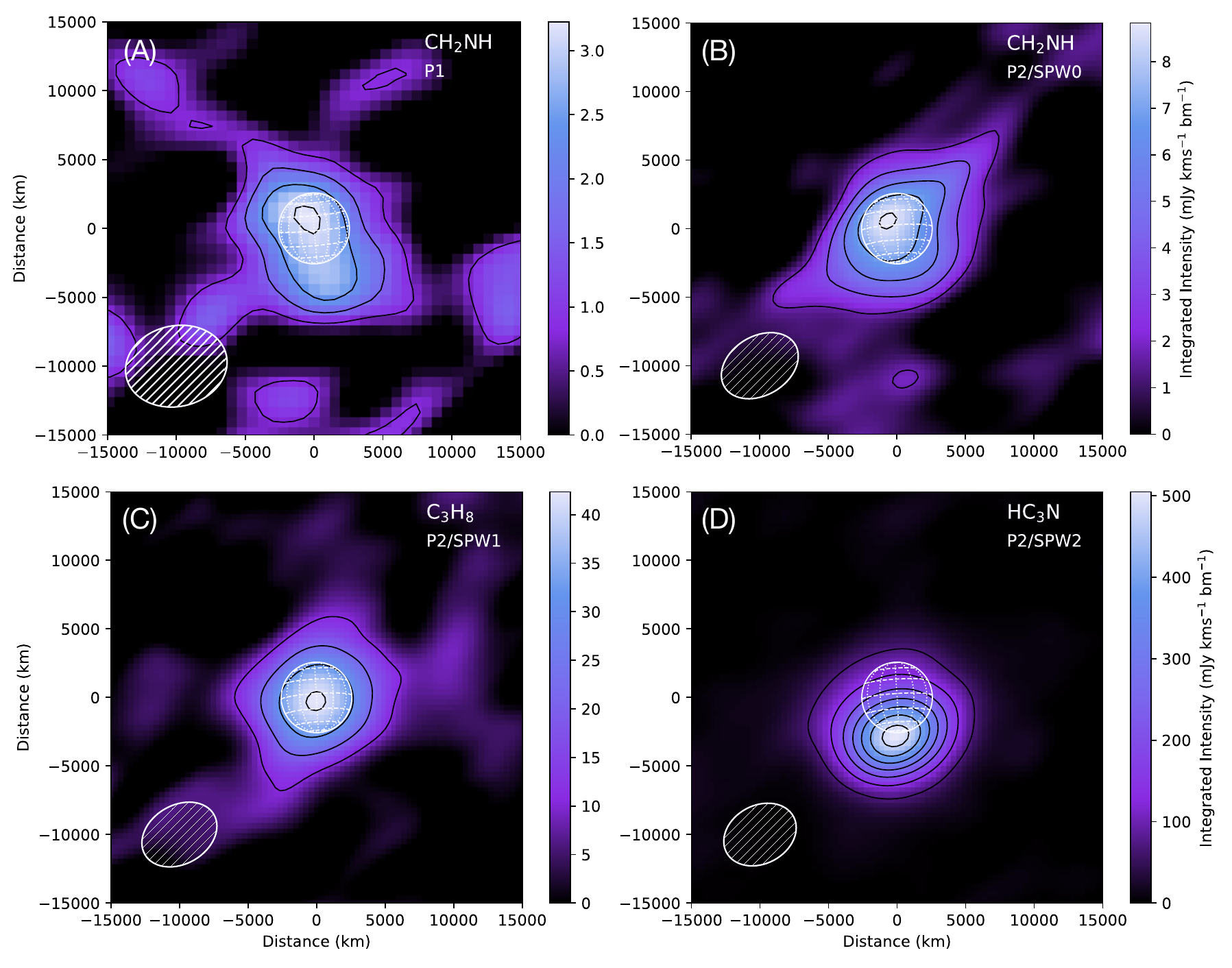} %{../Figures/Image_Maps/Maps_PBU.pdf}
  \caption{Integrated emission maps over resolved spectral features showing the
    distribution of CH$_2$NH in P1 and P2/SPW0 
    (panels A and B, respectively), C$_3$H$_8$ from P2/SPW1 (C) and
    HC$_3$N from P2/SPW2 (D) in
    Titan's atmosphere during 2024--2026. CH$_2$NH emission in panel B
    includes some underlying emission from the C$_3$H$_8$ band (see
    Figure \ref{fig:da_spec} middle panel); HC$_3$N emission in panel D was 
    measured from the $\nu_7=1$ transition (Table \ref{tab_spec};
    Figure \ref{fig:da_spec}, bottom panel). Contours
    correspond to 1-$\sigma$ (A), 2-$\sigma$ (B), 3-$\sigma$ (C), and 10-$\sigma$
    (D) levels. The ALMA beam is shown in the bottom left of each
    panel, shown here using ``natural'' baseline weighting to
    produce smoother emission distributions.}\label{fig:maps}
\end{figure}

 The list of detected spectral transitions, along with their
 upper-state energy levels ($E'$), are shown in Table
 \ref{tab_spec}. Following data calibration, Titan's continuum emission was
 subtracted from the resulting image cubes using the CASA task
 \texttt{imcontsub} to isolate and visualize the emission of trace
 gases in Titan's atmosphere. We then integrating the emission over all spectral
 line channels (using the CASA task \texttt{immoments}) to produce gas
 emission maps of CH$_2$NH, HC$_3$N, and C$_3$H$_8$, as shown in
 Figure \ref{fig:maps}. Although the integrated CH$_2$NH emission in
 both P1 and P2 (Figure \ref{fig:maps}A and B, respectively) does not
 exhibit preferential enhancement towards high latitudes, this may still indicate a lack of strong
concentration at Titan's winter pole (the South pole at the time of
observations). A polar enhancement has been observed in some
trace gases even in
observations with fairly coarse spatial resolution in the past
\citep{cordiner_14,cordiner_15,thelen_19b} and is evident in the
distribution of HC$_3$N produced from the 
observations analyzed here (Figure \ref{fig:maps}D). The distribution of
CH$_2$NH here is more similar to that of the concurrently detected
C$_3$H$_8$ (Figure \ref{fig:maps}C) and CH$_3$D emission previously
detected on Titan \citep{thelen_19b,thelen_24b}.

\section{Supplemental Radiative Transfer Details} \label{app:rt}
\subsection{Blended Feature Spectral Decomposition} \label{sec:decomp}
\begin{figure}[h]
 \centering
 \includegraphics[width=0.95\textwidth]{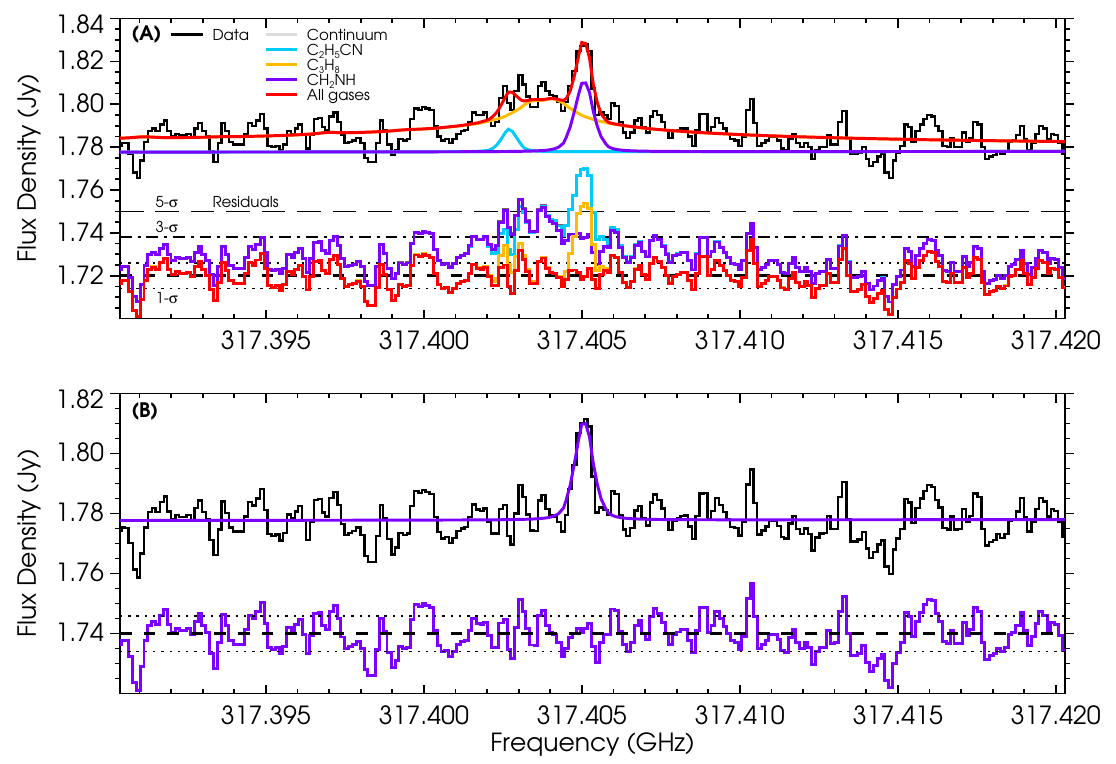} %{../Figures/spw0_decomp.pdf}
 \caption{ (A) ALMA spectrum from
   P2/SPW0 containing a blended feature comprising 
   C$_3$H$_8$, CH$_2$NH, and C$_2$H$_5$CN emission (black). Individual
 synthetic spectra are shown after adding in additional gases to the
 continuum model: C$_2$H$_5$CN (cyan), C$_3$H$_8$ (gold),
 CH$_2$NH (purple), and all three gases combined (red). Residual
 (data-model) spectra are shown in the
bottom panel (offset by +1.72 Jy for visibility), and 1-, 3-, and
5-$\sigma$ significance levels are shown for
reference (dotted, dash-dotted, and long dashed lines,
respectively). (B) Following the subtraction of the C$_2$H$_5$CN and
C$_3$H$_8$ emission (cyan and gold spectra in panel A), the CH$_2$NH
spectrum was isolated (black) and
fit separately (purple). The resulting best-fit spectrum corresponds
to the 2-point gradient profile as in Figures \ref{fig:combo_2}D. The
residual spectrum is adjusted by +1.74 Jy, with
1-$\sigma$ errors (dotted line) shown for reference.}\label{fig:decomp}
\end{figure}

While the CH$_2$NH spectra in P1 and P2/SPW2 were modeled without
significant interference from the surrounding spectral features, the CH$_2$NH
$J=5_{0,5}\rightarrow4_{0,4}$ transition in P2/SPW0 sits atop, and is
blended with, C$_3$H$_8$ transitions from the $J=20_8\rightarrow20_7$
rotational band. To ensure that the abundance profiles produced
through iterative retrievals of CH$_2$NH in
this spectrum were not influenced by the C$_3$H$_8$ emission (and, to
second order, emission from a proximal C$_2$H$_5$CN transition), we
first generated a forward model spectrum only consisting of emission from
potentially contaminating gases after fitting additional 
features found in adjacent spectral windows (e.g., the two C$_3$H$_8$
bands in P2/SPW1 -- Figure \ref{fig:combo_1}B). We subsequently
subtracted these spectra from the P2/SPW0 data
before retrieving the final CH$_2$NH profiles. The decomposition of
the full spectrum (as shown in Figure \ref{fig:da_spec}, middle panel)
is shown in Figure \ref{fig:decomp}.

\subsection{Full Photochemical Model Comparisons} \label{sec:pc}
\begin{figure}[h]
  \centering
  \includegraphics[width=0.95\textwidth]{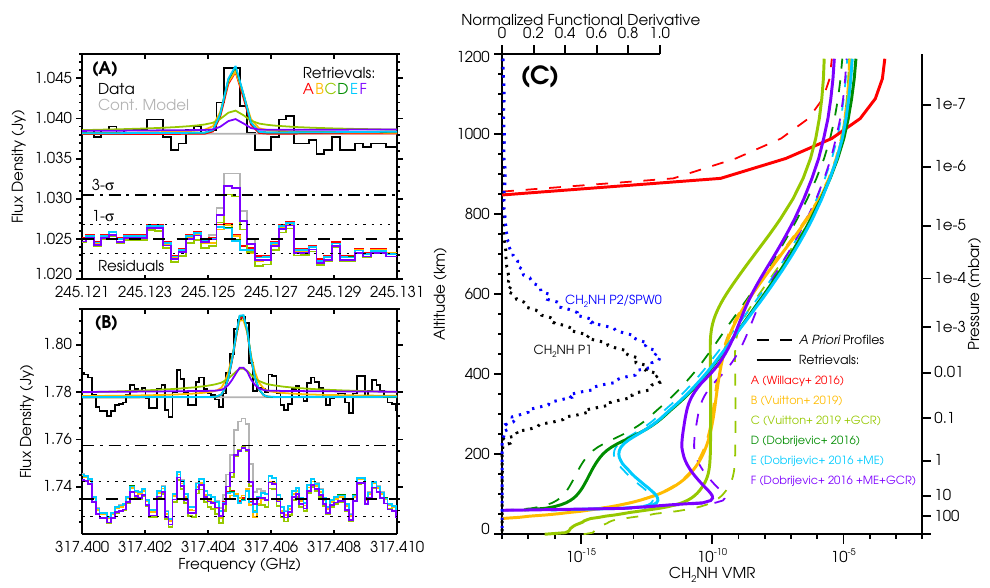} %{../Figures/ch2nh_mod_spec_comp_2_cfs}
  \caption{Comparisons of disk-averaged ALMA CH$_2$NH spectra with
    retrieved spectra initialized from photochemical model
    predictions. (A) ALMA data from P1 (black) compared to a model of
    Titan's continuum emission (gray) and various scaled CH$_2$NH
    models (colored spectra). These correspond to the model
    predictions from \citet{willacy_16} (model A, red line);
    \citet{vuitton_19} without (model B, gold) and with (model C,
    olive green) GCR
    chemistry; and models from \citet{dobrijevic_16} including
    ion-neutral chemistry only (model D, forest green), with magnetospheric
    electrons (model E, cyan), and also GCR (model F, purple). All
    models produce residual spectra (shown below + 1.025 Jy) with
    amplitudes largely \textless1-$\sigma$ (dotted line) except models with GCR
    chemistry included (models C, F). (B) Same as panel A but for
    spectrum from P2/SPW0 following the removal of C$_3$H$_8$ and
    C$_2$H$_5$CN emission (see Figure \ref{fig:decomp}). Residual
    spectra are adjusted for clarity (+1.735 Jy). (C) The input (dashed) and retrieved
    (solid) vertical profiles corresponding to the spectral fits in
    panels A and B; retrieved profiles were produced through the
    weighted average of all three CH$_2$NH spectral fits. The
    normalized functional derivatives of 
    the CH$_2$NH P1 and P2/SPW0 spectra as a function of
    CH$_2$NH abundance and altitude are shown (dotted black and blue lines,
    respectively), with the corresponding scale along the upper x-axis.}\label{fig:spec_comp}
\end{figure}

The full set of synthetic
spectra for all scaled photochemical models are shown in
Figure \ref{fig:spec_comp} compared to the CH$_2$NH P1 and
P2/SPW0 data, alongside the corresponding retrieved vertical
profiles. The altitude ranges where the CH$_2$NH P1 and P2/SPW0
radiances are most sensitive to changes in CH$_2$NH abundance (as determined by
the functional derivatives -- see \citet{irwin_08} for additional details) are also shown in Figure
\ref{fig:spec_comp}.

\end{appendices}

\pagebreak


\begin{thebibliography}{}
\expandafter\ifx\csname natexlab\endcsname\relax\def\natexlab#1{#1}\fi
\providecommand{\url}[1]{\href{#1}{#1}}
\providecommand{\dodoi}[1]{doi:~\href{http://doi.org/#1}{\nolinkurl{#1}}}
\providecommand{\doeprint}[1]{\href{http://ascl.net/#1}{\nolinkurl{http://ascl.net/#1}}}
\providecommand{\doarXiv}[1]{\href{https://arxiv.org/abs/#1}{\nolinkurl{https://arxiv.org/abs/#1}}}

\bibitem[{{Altwegg} {et~al.}(2016){Altwegg}, {Balsiger}, {Bar-Nun},
  {Berthelier}, {Bieler}, {Bochsler}, {Briois}, {Calmonte}, {Combi}, {Cottin},
  {De Keyser}, {Dhooghe}, {Fiethe}, {Fuselier}, {Gasc}, {Gombosi}, {Hansen},
  {Haessig}, {Ja ckel}, {Kopp}, {Korth}, {Le Roy}, {Mall}, {Marty}, {Mousis},
  {Owen}, {Reme}, {Rubin}, {Semon}, {Tzou}, {Waite}, \& {Wurz}}]{altwegg_16}
{Altwegg}, K., {Balsiger}, H., {Bar-Nun}, A., {et~al.} 2016, Science Advances,
  2, e1600285, \dodoi{10.1126/sciadv.1600285}

\bibitem[{{Barnes} {et~al.}(2021){Barnes}, {Turtle}, {Trainer}, {Lorenz},
  {MacKenzie}, {Brinckerhoff}, {Cable}, {Ernst}, {Freissinet}, {Hand}, {Hayes},
  {H{\"o}rst}, {Johnson}, {Karkoschka}, {Lawrence}, {Le Gall}, {Lora}, {McKay},
  {Miller}, {Murchie}, {Neish}, {Newman}, {N{\'u}{\~n}ez}, {Panning},
  {Parsons}, {Peplowski}, {Quick}, {Radebaugh}, {Rafkin}, {Shiraishi},
  {Soderblom}, {Sotzen}, {Stickle}, {Stofan}, {Szopa}, {Tokano}, {Wagner},
  {Wilson}, {Yingst}, {Zacny}, \& {St{\"a}hler}}]{barnes_21}
{Barnes}, J.~W., {Turtle}, E.~P., {Trainer}, M.~G., {et~al.} 2021, PSJ, 2, 130,
  \dodoi{10.3847/PSJ/abfdcf}

\bibitem[{{Belloche} {et~al.}(2008){Belloche}, {Menten}, {Comito},
  {M{\"u}ller}, {Schilke}, {Ott}, {Thorwirth}, \& {Hieret}}]{belloche_08}
{Belloche}, A., {Menten}, K.~M., {Comito}, C., {et~al.} 2008, A$\&$A, 482, 179,
  \dodoi{10.1051/0004-6361:20079203}

\bibitem[{Borysow(1991)}]{borysow_91}
Borysow, A. 1991, Icarus, 92, 273

\bibitem[{Borysow \& Frommhold(1986a)}]{borysow_86a}
Borysow, A., \& Frommhold, L. 1986a, ApJ, 311, 1043

\bibitem[{Borysow \& Frommhold(1986b)}]{borysow_86b}
---. 1986b, ApJ, 303, 495

\bibitem[{Borysow \& Frommhold(1986c)}]{borysow_86c}
---. 1986c, ApJ, 304, 849

\bibitem[{Borysow \& Frommhold(1987)}]{borysow_87}
---. 1987, ApJ, 318, 940

\bibitem[{Borysow \& Tang(1993)}]{borysow_93}
Borysow, A., \& Tang, C. 1993, Icarus, 105, 175

\bibitem[{{Briggs}(1995)}]{briggs_95}
{Briggs}, D.~S. 1995, in Bulletin of the American Astronomical Society,
  Vol.~27, American Astronomical Society Meeting Abstracts, 1444

\bibitem[{{Cerceau} {et~al.}(1985){Cerceau}, {Raulin}, {Courtin}, \&
  {Gautier}}]{cerceau_85}
{Cerceau}, F., {Raulin}, F., {Courtin}, R., \& {Gautier}, D. 1985, Icarus, 62,
  207, \dodoi{10.1016/0019-1035(85)90118-6}

\bibitem[{{Coates} {et~al.}(2007){Coates}, {Crary}, {Lewis}, {Young}, {Waite},
  \& {Sittler}}]{coates_07}
{Coates}, A.~J., {Crary}, F.~J., {Lewis}, G.~R., {et~al.} 2007, GRL, 34,
  L22103, \dodoi{10.1029/2007GL030978}

\bibitem[{Cordiner {et~al.}(2014)Cordiner, Nixon, Teanby, Irwin, Serigano,
  Charnley, Milam, Mumma, Lis, Villanueva, Paganini, Kuan, \&
  Remijan}]{cordiner_14}
Cordiner, M.~A., Nixon, C.~A., Teanby, N.~A., {et~al.} 2014, ApJ, 795, L30,
  \dodoi{10.1088/2041-8205/795/2/L30}

\bibitem[{Cordiner {et~al.}(2015)Cordiner, Palmer, Nixon, Irwin, Teanby,
  Charnley, Mumma, Kisiel, Serigano, Kuan, Chuang, \& Wang}]{cordiner_15}
Cordiner, M.~A., Palmer, M.~Y., Nixon, C.~A., {et~al.} 2015, ApJ, 800, L14,
  \dodoi{10.1088/2041-8205/800/1/L14}

\bibitem[{{Coustenis} {et~al.}(1998){Coustenis}, {Salama}, {Lellouch},
  {Encrenaz}, {Bjoraker}, {Samuelson}, {de Graauw}, {Feuchtgruber}, \&
  {Kessler}}]{coustenis_98}
{Coustenis}, A., {Salama}, A., {Lellouch}, E., {et~al.} 1998, A$\&$A, 336, L85

\bibitem[{{Cravens} {et~al.}(2008){Cravens}, {Robertson}, {Ledvina},
  {Mitchell}, {Krimigis}, \& {Waite}}]{cravens_08}
{Cravens}, T.~E., {Robertson}, I.~P., {Ledvina}, S.~A., {et~al.} 2008, GRL, 35,
  L03103, \dodoi{10.1029/2007GL032451}

\bibitem[{{Cravens} {et~al.}(2005){Cravens}, {Robertson}, {Clark}, {Wahlund},
  {Waite}, {Ledvina}, {Niemann}, {Yelle}, {Kasprzak}, {Luhmann}, {McNutt},
  {Ip}, {De La Haye}, {M{\"u}ller-Wodarg}, {Young}, \& {Coates}}]{cravens_05}
{Cravens}, T.~E., {Robertson}, I.~P., {Clark}, J., {et~al.} 2005, GRL, 32,
  L12108, \dodoi{10.1029/2005GL023249}

\bibitem[{Cui {et~al.}(2009a)Cui, Yelle, Vuitton, Waite~Jr., Kasprzak, Gell,
  Niemann, M{\"u}ller-Wodarg, Borggren, Fletcher, Patrick, Raaen, \&
  Magee}]{cui_09a}
Cui, J., Yelle, R.~V., Vuitton, V., {et~al.} 2009a, Icarus, 200, 581

\bibitem[{{Cui} {et~al.}(2009b){Cui}, {Galand}, {Yelle}, {Vuitton}, {Wahlund},
  {Lavvas}, {M{\"u}ller-Wodarg}, {Cravens}, {Kasprzak}, \& {Waite}}]{cui_09b}
{Cui}, J., {Galand}, M., {Yelle}, R.~V., {et~al.} 2009b, Journal of Geophysical
  Research (Space Physics), 114, A06310, \dodoi{10.1029/2009JA014228}

\bibitem[{{Danger} {et~al.}(2011){Danger}, {Borget}, {Chomat}, {Duvernay},
  {Theul{\'e}}, {Guillemin}, {Le Sergeant D'Hendecourt}, \&
  {Chiavassa}}]{danger_11}
{Danger}, G., {Borget}, F., {Chomat}, M., {et~al.} 2011, A$\&$A, 535, A47,
  \dodoi{10.1051/0004-6361/201117602}

\bibitem[{{Dickens} {et~al.}(1997){Dickens}, {Irvine}, {DeVries}, \&
  {Ohishi}}]{dickens_97}
{Dickens}, J.~E., {Irvine}, W.~M., {DeVries}, C.~H., \& {Ohishi}, M. 1997, ApJ,
  479, 307, \dodoi{10.1086/303884}

\bibitem[{{Dobrijevic} {et~al.}(2016){Dobrijevic}, {Loison}, {Hickson}, \&
  {Gronoff}}]{dobrijevic_16}
{Dobrijevic}, M., {Loison}, J.~C., {Hickson}, K.~M., \& {Gronoff}, G. 2016,
  Icarus, 268, 313, \dodoi{10.1016/j.icarus.2015.12.045}

\bibitem[{{Dore} {et~al.}(2012){Dore}, {Bizzocchi}, \& {Degli
  Esposti}}]{dore_12}
{Dore}, L., {Bizzocchi}, L., \& {Degli Esposti}, C. 2012, A$\&$A, 544, A19,
  \dodoi{10.1051/0004-6361/201219674}

\bibitem[{{Endres} {et~al.}(2016){Endres}, {Schlemmer}, {Schilke}, {Stutzki},
  \& {M{\"u}ller}}]{endres_16}
{Endres}, C.~P., {Schlemmer}, S., {Schilke}, P., {Stutzki}, J., \&
  {M{\"u}ller}, H. S.~P. 2016, Journal of Molecular Spectroscopy, 327, 95,
  \dodoi{10.1016/j.jms.2016.03.005}

\bibitem[{{Glavin} {et~al.}(2025){Glavin}, {Dworkin}, {Alexander}, {Aponte},
  {Baczynski}, {Barnes}, {Bechtel}, {Berger}, {Burton}, {Caselli}, {Chung},
  {Clemett}, {Cody}, {Dominguez}, {Elsila}, {Farnsworth}, {Foustoukos},
  {Freeman}, {Furukawa}, {Gainsforth}, {Graham}, {Grassi}, {Giuliano},
  {Hamilton}, {Haenecour}, {Heck}, {Hofmann}, {House}, {Huang}, {Kaplan},
  {Keller}, {Kim}, {Koga}, {Liss}, {McLain}, {Marcus}, {Matney}, {McCoy},
  {McIntosh}, {Mojarro}, {Naraoka}, {Nguyen}, {Nuevo}, {Nuth}, {Oba}, {Parker},
  {Peretyazhko}, {Sandford}, {Santos}, {Schmitt-Kopplin}, {Seguin}, {Simkus},
  {Shahid}, {Takano}, {Thomas-Keprta}, {Tripathi}, {Weiss}, {Zheng}, {Lunning},
  {Righter}, {Connolly}, \& {Lauretta}}]{glavin_25}
{Glavin}, D.~P., {Dworkin}, J.~P., {Alexander}, C. M.~O., {et~al.} 2025, Nature
  Astronomy, 9, 199, \dodoi{10.1038/s41550-024-02472-9}

\bibitem[{{Hanel} {et~al.}(1981){Hanel}, {Conrath}, {Flasar}, {Kunde},
  {Maguire}, {Pearl}, {Pirraglia}, {Samuelson}, {Herath}, {Allison},
  {Cruikshank}, {Gautier}, {Gierasch}, {Horn}, {Koppany}, \&
  {Ponnamperuma}}]{hanel_81}
{Hanel}, R., {Conrath}, B., {Flasar}, F.~M., {et~al.} 1981, Science, 212, 192,
  \dodoi{10.1126/science.212.4491.192}

\bibitem[{{Holtom} {et~al.}(2005){Holtom}, {Bennett}, {Osamura}, {Mason}, \&
  {Kaiser}}]{holtom_05}
{Holtom}, P.~D., {Bennett}, C.~J., {Osamura}, Y., {Mason}, N.~J., \& {Kaiser},
  R.~I. 2005, ApJ, 626, 940, \dodoi{10.1086/430106}

\bibitem[{{H{\"o}rst}(2017)}]{horst_17}
{H{\"o}rst}, S.~M. 2017, Journal of Geophysical Research (Planets), 122, 432,
  \dodoi{10.1002/2016JE005240}

\bibitem[{H{\"o}rst {et~al.}(2008)H{\"o}rst, Vuitton, \& Yelle}]{horst_08}
H{\"o}rst, S.~M., Vuitton, V., \& Yelle, R.~V. 2008, J. Geophys. Res.: Planets,
  113, \dodoi{10.1029/2008JE003135}

\bibitem[{{H{\"o}rst} {et~al.}(2012){H{\"o}rst}, {Yelle}, {Buch}, {Carrasco},
  {Cernogora}, {Dutuit}, {Quirico}, {Sciamma-O'Brien}, {Smith}, {Somogyi},
  {Szopa}, {Thissen}, \& {Vuitton}}]{horst_12}
{H{\"o}rst}, S.~M., {Yelle}, R.~V., {Buch}, A., {et~al.} 2012, Astrobiology,
  12, 809, \dodoi{10.1089/ast.2011.0623}

\bibitem[{Irwin {et~al.}(2008)Irwin, Teanby, de~Kok, Fletcher, Howett, Tsang,
  Wilson, Calcutt, Nixon, \& Parrish}]{irwin_08}
Irwin, P. G.~J., Teanby, N.~A., de~Kok, R., {et~al.} 2008, J. Quant. Spec.
  Radiat. Transf., 109, 1136, \dodoi{10.1016/j.jqsrt.2007.11.006}

\bibitem[{{Jaeger}(2008)}]{jaeger_08}
{Jaeger}, S. 2008, in Astronomical Society of the Pacific Conference Series,
  Vol. 394, Astronomical Data Analysis Software and Systems XVII, ed. R.~W.
  {Argyle}, P.~S. {Bunclark}, \& J.~R. {Lewis}, 623

\bibitem[{{Kim} \& {Kaiser}(2011)}]{kim_11}
{Kim}, Y.~S., \& {Kaiser}, R.~I. 2011, ApJ, 729, 68,
  \dodoi{10.1088/0004-637X/729/1/68}

\bibitem[{Kisiel {et~al.}(2020)Kisiel, Nixon, Cordiner, Thelen, \&
  Charnley}]{kisiel_20}
Kisiel, Z., Nixon, C.~A., Cordiner, M.~A., Thelen, A.~E., \& Charnley, S.~B.
  2020, {Journal of Molecular Spectroscopy}, 372, 111324,
  \dodoi{10.1016/j.jms.2020.111324}

\bibitem[{{Krasnopolsky}(2009)}]{krasnopolsky_09}
{Krasnopolsky}, V.~A. 2009, Icarus, 201, 226,
  \dodoi{10.1016/j.icarus.2008.12.038}

\bibitem[{{Krause} \& {Sutter}(1989)}]{krause_89}
{Krause}, H., \& {Sutter}, D.~H. 1989, Zeitschrift Naturforschung Teil A, 44,
  1063, \dodoi{10.1515/zna-1989-1106}

\bibitem[{{Kunde} {et~al.}(1981){Kunde}, {Aikin}, {Hanel}, {Jennings},
  {Maguire}, \& {Samuelson}}]{kunde_81}
{Kunde}, V.~G., {Aikin}, A.~C., {Hanel}, R.~A., {et~al.} 1981, Nature, 292,
  686, \dodoi{10.1038/292686a0}

\bibitem[{{Kvenvolden} {et~al.}(1970){Kvenvolden}, {Lawless}, {Pering},
  {Peterson}, {Flores}, \& {Ponnamperuma}}]{kvenvolden_70}
{Kvenvolden}, K., {Lawless}, J., {Pering}, K., {et~al.} 1970, Nature, 228, 923,
  \dodoi{10.1038/228923a0}

\bibitem[{{Lai} {et~al.}(2017){Lai}, {Cordiner}, {Nixon}, {Achterberg},
  {Molter}, {Teanby}, {Palmer}, {Charnley}, {Lindberg}, {Kisiel}, {Mumma}, \&
  {Irwin}}]{lai_17}
{Lai}, J.~C.-Y., {Cordiner}, M.~A., {Nixon}, C.~A., {et~al.} 2017, AJ, 154,
  206, \dodoi{10.3847/1538-3881/aa8eef}

\bibitem[{Lavvas {et~al.}(2008)Lavvas, Coustenis, \& Vardavas}]{lavvas_08a}
Lavvas, P.~P., Coustenis, A., \& Vardavas, I.~M. 2008, Planet. Space Sci., 56,
  27

\bibitem[{{Lin} {et~al.}(2026){Lin}, {Spezzano}, {Sipil{\"a}}, {Pineda}, \&
  {Caselli}}]{lin_26}
{Lin}, Y., {Spezzano}, S., {Sipil{\"a}}, O., {Pineda}, J.~E., \& {Caselli}, P.
  2026, ApJL, 996, L32, \dodoi{10.3847/2041-8213/ae2c76}

\bibitem[{Loison {et~al.}(2015)Loison, H{\'e}brard, Dobrijevic, Hickson,
  Caralp, Hue, Gronoff, Venot, \& B{\'e}nilan}]{loison_15}
Loison, J.~C., H{\'e}brard, E., Dobrijevic, M., {et~al.} 2015, Icarus, 247,
  218, \dodoi{10.1016/j.icarus.2014.09.039}

\bibitem[{{Lopes} {et~al.}(2020){Lopes}, {Malaska}, {Schoenfeld},
  {Solomonidou}, {Birch}, {Florence}, {Hayes}, {Williams}, {Radebaugh},
  {Verlander}, {Turtle}, {Le Gall}, \& {Wall}}]{lopes_20}
{Lopes}, R.~M.~C., {Malaska}, M.~J., {Schoenfeld}, A.~M., {et~al.} 2020, Nature
  Astronomy, 4, 228, \dodoi{10.1038/s41550-019-0917-6}

\bibitem[{{Lopes} {et~al.}(2026){Lopes}, {Malaska}, {Vance}, {Schoenfeld},
  {Solomonidou}, {Hodyss}, {Meyer-Dombard}, {Fagents}, \& {Titan NAI
  Team}}]{lopes_26}
{Lopes}, R.~M.~C., {Malaska}, M.~J., {Vance}, S.~D., {et~al.} 2026, in LPI
  Contributions, Vol. 3160, 57th Lunar and Planetary Science Conference, 1313

\bibitem[{Marten {et~al.}(2002)Marten, Hidayat, Biraud, \& Moreno}]{marten_02}
Marten, A., Hidayat, T., Biraud, Y., \& Moreno, R. 2002, Icarus, 158, 532

\bibitem[{{McGuire}(2022)}]{mcguire_22}
{McGuire}, B.~A. 2022, ApJ Supp. Series, 259, 30,
  \dodoi{10.3847/1538-4365/ac2a48}

\bibitem[{Molter {et~al.}(2016)Molter, Nixon, Cordiner, Serigano, Irwin,
  Teanby, Charnley, \& Lindberg}]{molter_16}
Molter, E.~M., Nixon, C.~A., Cordiner, M.~A., {et~al.} 2016, AJ, 152, 1,
  \dodoi{10.3847/0004-6256/152/2/42}

\bibitem[{{Moreno} {et~al.}(2011){Moreno}, {Lellouch}, {Lara}, {Courtin},
  {Bockel{\'e}e-Morvan}, {Hartogh}, {Rengel}, {Biver}, {Banaszkiewicz}, \&
  {Gonz{\'a}lez}}]{moreno_11}
{Moreno}, R., {Lellouch}, E., {Lara}, L.~M., {et~al.} 2011, A$\&$A, 536, L12,
  \dodoi{10.1051/0004-6361/201118189}

\bibitem[{{M{\"u}ller} {et~al.}(2005){M{\"u}ller}, {Schl{\"o}der}, {Stutzki},
  \& {Winnewisser}}]{muller_05}
{M{\"u}ller}, H.~S.~P., {Schl{\"o}der}, F., {Stutzki}, J., \& {Winnewisser}, G.
  2005, Journal of Molecular Structure, 742, 215,
  \dodoi{10.1016/j.molstruc.2005.01.027}

\bibitem[{{Muller} {et~al.}(2011){Muller}, {Beelen}, {Gu{\'e}lin}, {Aalto},
  {Black}, {Combes}, {Curran}, {Theule}, \& {Longmore}}]{muller_11}
{Muller}, S., {Beelen}, A., {Gu{\'e}lin}, M., {et~al.} 2011, A$\&$A, 535, A103,
  \dodoi{10.1051/0004-6361/201117096}

\bibitem[{{Neish} {et~al.}(2024){Neish}, {Malaska}, {Sotin}, {Lopes}, {Nixon},
  {Affholder}, {Chatain}, {Cockell}, {Farnsworth}, {Higgins}, {Miller}, \&
  {Soderlund}}]{neish_24}
{Neish}, C., {Malaska}, M.~J., {Sotin}, C., {et~al.} 2024, Astrobiology, 24,
  177, \dodoi{10.1089/ast.2023.0055}

\bibitem[{{Neish} {et~al.}(2010){Neish}, {Somogyi}, \& {Smith}}]{neish_10}
{Neish}, C.~D., {Somogyi}, {\'A}., \& {Smith}, M.~A. 2010, Astrobiology, 10,
  337, \dodoi{10.1089/ast.2009.0402}

\bibitem[{{Neish} {et~al.}(2015){Neish}, {Barnes}, {Sotin}, {MacKenzie},
  {Soderblom}, {Le Mou{\'e}lic}, {Kirk}, {Stiles}, {Malaska}, {Le Gall},
  {Brown}, {Baines}, {Buratti}, {Clark}, \& {Nicholson}}]{neish_15}
{Neish}, C.~D., {Barnes}, J.~W., {Sotin}, C., {et~al.} 2015, GRL, 42, 3746,
  \dodoi{10.1002/2015GL063824}

\bibitem[{{Neish} {et~al.}(2018){Neish}, {Lorenz}, {Turtle}, {Barnes},
  {Trainer}, {Stiles}, {Kirk}, {Hibbitts}, \& {Malaska}}]{neish_18}
{Neish}, C.~D., {Lorenz}, R.~D., {Turtle}, E.~P., {et~al.} 2018, Astrobiology,
  18, 571, \dodoi{10.1089/ast.2017.1758}

\bibitem[{Niemann {et~al.}(2005)Niemann, Atreya, Bauer, Carignan, Demick,
  Frost, Gautier, Haberman, Harpold, Hunten, Israel, Lunine, Kasprzak, Owen,
  Paulkovich, Raulin, Raaen, \& Way}]{niemann_05}
Niemann, H.~B., Atreya, S.~K., Bauer, S.~J., {et~al.} 2005, Nature, 438, 779

\bibitem[{Niemann {et~al.}(2010)Niemann, Atreya, Demick, Gautier, Haberman,
  Harpold, Kasprzak, Lunine, Owen, \& Raulin}]{niemann_10}
Niemann, H.~B., Atreya, S.~K., Demick, J.~E., {et~al.} 2010, JGR, 115,
  \dodoi{10.1029/2010JE003659}

\bibitem[{{Nixon}(2024)}]{nixon_24}
{Nixon}, C.~A. 2024, ACS Earth and Space Chemistry, 8, 406,
  \dodoi{10.1021/acsearthspacechem.2c00041}

\bibitem[{Nixon {et~al.}(2020)Nixon, Thelen, Cordiner, Kisiel, Charnley,
  Molter, Serigano, Irwin, Teanby, \& Kuan}]{nixon_20}
Nixon, C.~A., Thelen, A.~E., Cordiner, M.~A., {et~al.} 2020, {AJ}, 160, 205

\bibitem[{{Or{\'o}}(1961)}]{oro_61b}
{Or{\'o}}, J. 1961, Nature, 191, 1193, \dodoi{10.1038/1911193a0}

\bibitem[{{Or{\'o}} \& {Kamat}(1961)}]{oro_61a}
{Or{\'o}}, J., \& {Kamat}, S.~S. 1961, Nature, 190, 442,
  \dodoi{10.1038/190442a0}

\bibitem[{Palmer {et~al.}(2017)Palmer, Cordiner, Nixon, Charnley, Teanby,
  Kisiel, Irwin, \& Mumma}]{palmer_17}
Palmer, M.~Y., Cordiner, M.~A., Nixon, C.~A., {et~al.} 2017, Sci. Adv., 3,
  \dodoi{10.1126/sciadv.1700022}

\bibitem[{{Peltzer} {et~al.}(1984){Peltzer}, {Bada}, {Schlesinger}, \&
  {Miller}}]{peltzer_84}
{Peltzer}, E.~T., {Bada}, J.~L., {Schlesinger}, G., \& {Miller}, S.~L. 1984,
  Advances in Space Research, 4, 69, \dodoi{10.1016/0273-1177(84)90546-5}

\bibitem[{{Petricca} {et~al.}(2025){Petricca}, {Vance}, {Parisi}, {Buccino},
  {Cascioli}, {Castillo-Rogez}, {Downey}, {Nimmo}, {Tobie}, {Journaux},
  {Magnanini}, {Jones}, {Panning}, {Bagheri}, {Genova}, \&
  {Lunine}}]{petricca_25}
{Petricca}, F., {Vance}, S.~D., {Parisi}, M., {et~al.} 2025, Nature, 648, 556,
  \dodoi{10.1038/s41586-025-09818-x}

\bibitem[{{Poch} {et~al.}(2012){Poch}, {Coll}, {Buch}, {Ram{\'\i}rez}, \&
  {Raulin}}]{poch_12}
{Poch}, O., {Coll}, P., {Buch}, A., {Ram{\'\i}rez}, S.~I., \& {Raulin}, F.
  2012, P$\&$SS, 61, 114, \dodoi{10.1016/j.pss.2011.04.009}

\bibitem[{{Raulin} \& {Owen}(2002)}]{raulin_02}
{Raulin}, F., \& {Owen}, T. 2002, Space Sci. Rev., 104, 377,
  \dodoi{10.1023/A:1023636623006}

\bibitem[{{Roy} {et~al.}(2007){Roy}, {Najafian}, \& {von Ragu{\'e}
  Schleyer}}]{roy_07}
{Roy}, D., {Najafian}, K., \& {von Ragu{\'e} Schleyer}, P. 2007, Proceedings of
  the National Academy of Science, 104, 17272, \dodoi{10.1073/pnas.0708434104}

\bibitem[{{Sagan} \& {Khare}(1979)}]{sagan_79}
{Sagan}, C., \& {Khare}, B.~N. 1979, Nature, 277, 102, \dodoi{10.1038/277102a0}

\bibitem[{{Salter} {et~al.}(2008){Salter}, {Ghosh}, {Catinella}, {Lebron},
  {Lerner}, {Minchin}, \& {Momjian}}]{salter_08}
{Salter}, C.~J., {Ghosh}, T., {Catinella}, B., {et~al.} 2008, AJ, 136, 389,
  \dodoi{10.1088/0004-6256/136/1/389}

\bibitem[{{Samuelson} {et~al.}(1983){Samuelson}, {Maguire}, {Hanel},
  {Jennings}, {Yung}, \& {Aikin}}]{samuelson_83b}
{Samuelson}, R.~E., {Maguire}, W.~C., {Hanel}, R.~A., {et~al.} 1983, JGR, 88,
  8709, \dodoi{10.1029/JA088iA11p08709}

\bibitem[{{Schinder} {et~al.}(2020){Schinder}, {Flasar}, {Marouf}, {French},
  {Anabtawi}, {Barbinis}, {Fleischman}, \& {Achterberg}}]{schinder_20}
{Schinder}, P.~J., {Flasar}, F.~M., {Marouf}, E.~A., {et~al.} 2020, Icarus,
  345, 113720, \dodoi{10.1016/j.icarus.2020.113720}

\bibitem[{Serigano {et~al.}(2016)Serigano, Nixon, Cordiner, Irwin, Teanby,
  Charnley, \& Lindberg}]{serigano_16}
Serigano, J., Nixon, C.~A., Cordiner, M.~A., {et~al.} 2016, ApJ, 821, L8,
  \dodoi{10.3847/2041-8205/821/1/L8}

\bibitem[{{Solomonidou} {et~al.}(2020){Solomonidou}, {Neish}, {Coustenis},
  {Malaska}, {Le Gall}, {Lopes}, {Werynski}, {Markonis}, {Lawrence},
  {Altobelli}, {Witasse}, {Schoenfeld}, {Matsoukas}, {Baziotis}, \&
  {Drossart}}]{solomonidou_20}
{Solomonidou}, A., {Neish}, C., {Coustenis}, A., {et~al.} 2020, A$\&$A, 641,
  A16, \dodoi{10.1051/0004-6361/202037866}

\bibitem[{{Sotin} {et~al.}(2021){Sotin}, {Kalousov{\'a}}, \&
  {Tobie}}]{sotin_21}
{Sotin}, C., {Kalousov{\'a}}, K., \& {Tobie}, G. 2021, Annual Review of Earth
  and Planetary Sciences, 49, \dodoi{10.1146/annurev-earth-072920-052847}

\bibitem[{{Spezzano} {et~al.}(2022){Spezzano}, {Caselli}, {Sipil{\"a}}, \&
  {Bizzocchi}}]{spezzano_22}
{Spezzano}, S., {Caselli}, P., {Sipil{\"a}}, O., \& {Bizzocchi}, L. 2022,
  A$\&$A, 664, L2, \dodoi{10.1051/0004-6361/202244301}

\bibitem[{{Suzuki} {et~al.}(2016){Suzuki}, {Ohishi}, {Hirota}, {Saito},
  {Majumdar}, \& {Wakelam}}]{suzuki_16}
{Suzuki}, T., {Ohishi}, M., {Hirota}, T., {et~al.} 2016, ApJ, 825, 79,
  \dodoi{10.3847/0004-637X/825/1/79}

\bibitem[{Teanby {et~al.}(2009{\natexlab{a}})Teanby, Irwin, de~Kok, Jolly,
  B{\'e}zard, Nixon, \& Calcutt}]{teanby_09a}
Teanby, N.~A., Irwin, P. G.~J., de~Kok, R., {et~al.} 2009{\natexlab{a}},
  Icarus, 202, 620

\bibitem[{Teanby {et~al.}(2009{\natexlab{b}})Teanby, Irwin, de~Kok, \&
  Nixon}]{teanby_09b}
Teanby, N.~A., Irwin, P. G.~J., de~Kok, R., \& Nixon, C.~A. 2009{\natexlab{b}},
  Phil. Trans. R. Soc. A, 367, 697

\bibitem[{Teanby {et~al.}(2010)Teanby, Irwin, de~Kok, \& Nixon}]{teanby_10a}
---. 2010, ApJ, 724, L84, \dodoi{10.1088/2041-8205/724/1/L84}

\bibitem[{Teanby {et~al.}(2008)Teanby, de~Kok, Irwin, Osprey, Vinatier,
  Gierasch, Read, Flasar, Conrath, Achterberg, B{\'e}zard, Nixon, \&
  Calcutt}]{teanby_08b}
Teanby, N.~A., de~Kok, R., Irwin, P. G.~J., {et~al.} 2008, J. Geophys. Res.,
  113, 1

\bibitem[{Teanby {et~al.}(2013)Teanby, Irwin, Nixon, Courtin, Swinyard, Moreno,
  Lellouch, Rengel, \& Hartogh}]{teanby_13}
Teanby, N.~A., Irwin, P. G.~J., Nixon, C.~A., {et~al.} 2013, Planet. Space
  Sci., 75, 136, \dodoi{10.1016/j.pss.2012.11.008}

\bibitem[{Teanby {et~al.}(2018)Teanby, Cordiner, Nixon, Irwin, H{\"o}rst,
  Sylvestre, Serigano, Thelen, Richards, \& Charnley}]{teanby_18}
Teanby, N.~A., Cordiner, M.~A., Nixon, C.~A., {et~al.} 2018, AJ, 155, 251,
  \dodoi{10.3847/1538-3881/aac172}

\bibitem[{{Thelen} {et~al.}(2022){Thelen}, {Nixon}, {Cosentino}, {Cordiner},
  {Teanby}, {Newman}, {Irwin}, \& {Charnley}}]{thelen_22}
{Thelen}, A.~E., {Nixon}, C.~A., {Cosentino}, R.~G., {et~al.} 2022, PSJ, 3,
  146, \dodoi{10.3847/PSJ/ac7050}

\bibitem[{{Thelen} {et~al.}(2019{\natexlab{a}}){Thelen}, {Nixon}, {Charnley},
  {Irwin}, \& {Kisiel}}]{thelen_19b}
{Thelen}, A.~E., {Nixon}, C.~A.and~{Cordiner}, M.~A., {Charnley}, S.~B.,
  {Irwin}, P.~G.~J., \& {Kisiel}, Z. 2019{\natexlab{a}}, AJ, 157, 219,
  \dodoi{10.3847/1538-3881/ab19bb}

\bibitem[{{Thelen} {et~al.}(2018){Thelen}, {Nixon}, {Chanover}, {Molter},
  {Cordiner}, {Achterberg}, {Serigano}, {Irwin}, {Teanby}, \&
  {Charnley}}]{thelen_18}
{Thelen}, A.~E., {Nixon}, C.~A., {Chanover}, N.~J., {et~al.} 2018, Icarus, 307,
  380, \dodoi{10.1016/j.icarus.2017.10.042}

\bibitem[{{Thelen} {et~al.}(2019{\natexlab{b}}){Thelen}, {Nixon}, {Chanover},
  {Cordiner}, {Molter}, {Teanby}, {Irwin}, {Serigano}, \&
  {Charnley}}]{thelen_19a}
---. 2019{\natexlab{b}}, Icarus, 319, 417, \dodoi{10.1016/j.icarus.2018.09.023}

\bibitem[{{Thelen} {et~al.}(2020){Thelen}, {Cordiner}, {Nixon}, {Vuitton},
  {Kisiel}, {Charnley}, {Palmer}, {Teanby}, \& {Irwin}}]{thelen_20}
{Thelen}, A.~E., {Cordiner}, M.~A., {Nixon}, C.~A., {et~al.} 2020, ApJL, 903,
  L22, \dodoi{10.3847/2041-8213/abc1e1}

\bibitem[{{Thelen} {et~al.}(2024){Thelen}, {Nixon}, {Cordiner}, {Lellouch},
  {Vinatier}, {Teanby}, {Butler}, {Charnley}, {Cosentino}, {de Kleer}, {Irwin},
  {Gurwell}, {Kisiel}, \& {Moreno}}]{thelen_24b}
{Thelen}, A.~E., {Nixon}, C.~A., {Cordiner}, M.~A., {et~al.} 2024, PSJ, 5, 125,
  \dodoi{10.3847/PSJ/ad47bd}

\bibitem[{{Theule} {et~al.}(2011){Theule}, {Borget}, {Mispelaer}, {Danger},
  {Duvernay}, {Guillemin}, \& {Chiavassa}}]{theule_11}
{Theule}, P., {Borget}, F., {Mispelaer}, F., {et~al.} 2011, A$\&$A, 534, A64,
  \dodoi{10.1051/0004-6361/201117494}

\bibitem[{{Turtle} {et~al.}(2011){Turtle}, {Perry}, {Hayes}, {Lorenz},
  {Barnes}, {McEwen}, {West}, {Del Genio}, {Barbara}, {Lunine}, {Schaller},
  {Ray}, {Lopes}, \& {Stofan}}]{turtle_11}
{Turtle}, E.~P., {Perry}, J.~E., {Hayes}, A.~G., {et~al.} 2011, Science, 331,
  1414, \dodoi{10.1126/science.1201063}

\bibitem[{Vinatier {et~al.}(2007)Vinatier, B{\'e}zard, Fouchet, Teanby, de~Kok,
  Irwin, Conrath, Nixon, Romani, Flasar, \& Coustenis}]{vinatier_07a}
Vinatier, S., B{\'e}zard, B., Fouchet, T., {et~al.} 2007, Icarus, 188, 120

\bibitem[{Vinatier {et~al.}(2015)Vinatier, B{\'e}zard, Lebonnois, Teanby,
  Achterberg, Gorius, Mamoutkine, Guandique, Jolly, Jennings, \&
  Flasar}]{vinatier_15}
Vinatier, S., B{\'e}zard, B., Lebonnois, S., {et~al.} 2015, Icarus, 250, 95

\bibitem[{Vuitton {et~al.}(2006)Vuitton, Yelle, \& Anicich}]{vuitton_06}
Vuitton, V., Yelle, R.~V., \& Anicich, V.~G. 2006, ApJ, 647, L175

\bibitem[{{Vuitton} {et~al.}(2019){Vuitton}, {Yelle}, {Klippenstein},
  {H{\"o}rst}, \& {Lavvas}}]{vuitton_19}
{Vuitton}, V., {Yelle}, R.~V., {Klippenstein}, S.~J., {H{\"o}rst}, S.~M., \&
  {Lavvas}, P. 2019, Icarus, 324, 120, \dodoi{10.1016/j.icarus.2018.06.013}

\bibitem[{Vuitton {et~al.}(2007)Vuitton, Yelle, \& McEwan}]{vuitton_07}
Vuitton, V., Yelle, R.~V., \& McEwan, M.~J. 2007, Icarus, 191, 722,
  \dodoi{10.1016/j.icarus.2007.06.023}

\bibitem[{{Waite} {et~al.}(2005){Waite}, {Niemann}, {Yelle}, {Kasprzak},
  {Cravens}, {Luhmann}, {McNutt}, {Ip}, {Gell}, {De La Haye},
  {M{\"u}ller-Wordag}, {Magee}, {Borggren}, {Ledvina}, {Fletcher}, {Walter},
  {Miller}, {Scherer}, {Thorpe}, {Xu}, {Block}, \& {Arnett}}]{waite_05}
{Waite}, J.~H., {Niemann}, H., {Yelle}, R.~V., {et~al.} 2005, Science, 308,
  982, \dodoi{10.1126/science.1110652}

\bibitem[{{Willacy} {et~al.}(2016){Willacy}, {Allen}, \& {Yung}}]{willacy_16}
{Willacy}, K., {Allen}, M., \& {Yung}, Y. 2016, ApJ, 829, 79,
  \dodoi{10.3847/0004-637X/829/2/79}

\bibitem[{Wilson \& Atreya(2004)}]{wilson_04}
Wilson, E.~H., \& Atreya, S.~K. 2004, J. Geophys. Res., 109, 1

\bibitem[{Yelle {et~al.}(2010)Yelle, Vuitton, Lavvas, Klippenstein, Smith,
  H{\"o}rst, \& Cui}]{yelle_10}
Yelle, R.~V., Vuitton, V., Lavvas, P., {et~al.} 2010, Faraday Discuss., 147, 31

\bibitem[{Yung {et~al.}(1984)Yung, Allen, \& Pinto}]{yung_84}
Yung, Y.~L., Allen, M., \& Pinto, J.~P. 1984, ApJ SS, 55, 465

\end{thebibliography}
\end{document}